\documentclass[structabstract]{aa}  
\usepackage{graphicx}
\usepackage{natbib}
\bibpunct{(}{)}{;}{a}{}{,}

\usepackage{txfonts}
\begin{document}
   \title{Particle-in-cell simulation of a mildly relativistic collision 
    of an electron-ion plasma carrying a quasi-parallel magnetic field}

   \subtitle{Electron acceleration and magnetic field amplification at
   supernova shocks}

   \author{M. E. Dieckmann
          \inst{1}
          \and
           G. C. Murphy
          \inst{2}
          \and
           A. Meli
          \inst{3}
          \and
           L. O. C. Drury
          \inst{2}
          }

   \institute{Department of Science and Technology, 
              Linkoping University, 60174 Norrkoping, Sweden\\
              \email{Mark.E.Dieckmann@itn.liu.se}
                       \and
              Dublin Institute for Advanced Studies,
              31 Fitzwilliam Place, Dublin 2, Ireland 
         \and
              Center for Astroparticle Physics,
              Friedrich-Alexander-University Erlangen-Nuremberg,
              91058 Erlangen, Germany
             }
 \titlerunning{2D PIC simulation of an oblique shock}
    
   \date{Received September 15, 1996; accepted March 16, 1997}

\abstract
{Plasma processes close to supernova remnant shocks result in the
amplification of magnetic fields and in the acceleration of electrons, 
injecting them into the diffusive acceleration mechanism.}
{The acceleration of electrons and the magnetic field amplification by
the collision of two plasma clouds, each consisting of electrons and
ions, at a speed of 0.5c is investigated. A quasi-parallel guiding magnetic 
field, a cloud density ratio of 10 and a plasma temperature of 25 keV
are considered.}
{A relativistic and electromagnetic particle-in-cell simulation models 
the plasma in two spatial dimensions employing an ion-to-electron 
mass ratio of 400.}
{A quasi-planar shock forms at the front of the dense plasma cloud. It is 
mediated by a circularly left-hand polarized electromagnetic wave with an 
electric field component along the guiding magnetic field. Its propagation
direction is close to that of the guiding field and orthogonal to the 
collision boundary. It has a frequency too low to be determined during
the simulation time and a wavelength that equals several times the ion 
inertial length. These properties would be indicative of a dispersive 
Alfv\'en wave close to the ion cyclotron resonance frequency of the left-handed 
mode, known as the ion whistler, provided that the frequency is appropriate. 
However, it moves with the super-alfv\'enic plasma collision speed, suggesting 
that it is an Alfv\'en precursor or a nonlinear MHD wave such as 
a Short Large-Amplitude Magnetic Structure (SLAMS). The growth of the magnetic 
amplitude of this wave to values well in excess of those of the quasi-parallel 
guiding field and of the filamentation modes results in a quasi-perpendicular 
shock. We present evidence for the instability of this mode to a four 
wave interaction. The waves developing upstream of the dense cloud give rise 
to electron acceleration ahead of the collision boundary. Energy 
equipartition between the ions and the electrons is established at the shock 
and the electrons are accelerated to relativistic speeds.}
{The magnetic fields in the foreshock of supernova remnant shocks can be 
amplified substantially and electrons can be injected into the diffusive 
acceleration, if strongly magnetised plasma subshells are present in the 
foreshock, with velocities an order of magnitude faster than the main shell.}


   \maketitle
%

\section{Introduction}
Supernova remnants (SNRs) emanate energetic electromagnetic radiation,
which demonstrates the acceleration of electrons to ultrarelativistic speeds 
\citep{RelEl1,Radiation} and the generation or amplification of magnetic 
fields \citep{MagAmp1,MagAmp2}. The likely origin of the accelerated 
electrons and of the strong magnetic fields is the SNR shock
\citep{Marco1,Marco2}.

The nonrelativistic expansion speed of the main SNR blast shell 
\citep{Shockspeed1,Shockspeed2} and the weak magnetic field of the ambient 
medium \citep{MagAmp2}, into which this shell is expanding, are obstacles 
to the magnetic field amplification by plasma instabilities and to the
electron acceleration out of the thermal plasma distribution to moderately 
relativistic energies. Such an acceleration is needed for their injection 
\citep{Injection1,Injection2,Injection3,Gyrosurf} into the diffusive 
shock acceleration process (See \citet{Diffus}) so that they can cross 
the shock transition layer repeatedly. Electrostatic instabilities dominate 
for nonrelativistic flows in unmagnetized plasmas \citep{Bret,BRETAPJ} and 
they can neither accelerate the electrons to highly relativistic speeds 
\citep{Sircombe} nor amplify the magnetic fields.  

The electrons could be accelerated by plasma based charged particle 
accelerators \citep{PlAc}, by electron surfing acceleration 
\citep{Surfing1,Surfing2,Surfing3}, double layers \citep{DL2,DL1} or 
by processes that exploit a velocity shear in the plasma outflow 
\citep{Shear}. It has, however, not yet been demonstrated with 
multi-dimensional and self-consistent simulations that these 
mechanisms can indeed achieve the required electron acceleration and 
magnetic field amplification. The non-relativistic shocks, which are 
found between the main SNR blast shell and the ambient medium, can 
also probably not transfer significant energy from the ions to the 
electrons and accelerate the latter to relativistic speeds 
\citep{Shock4,Shock6,Shock5,Shock3,Sorasio,Shock2,Shock7}.

A viable acceleration mechanism may develop ahead of the main SNR shock, 
if we find subshells that outrun the main blast shell. These subshells may 
move faster than the typical peak speed of 0.2c of the main shell. An 
expansion speed as high as 0.9c may have been observed for a subshell 
ejected by the supernova SN1998bw \citep{Shockspeed1}. Most supernovae 
are less violent and their subshells are probably slower. The 
density of the subshell plasma is well below that of the main blast shell 
and its dynamics will be influenced to a larger extent by the upstream 
magnetic field than the dynamics of the latter. This is true in particular, 
if the upstream magnetic field has been pre-amplified by the cosmic rays 
\citep{Winske,MagAmp3,MagAmp4,Pohl,Riq}. A fast magnetized shock would 
form in the foreshock of the main SNR shock that can result in a 
stronger magnetic field amplification and electron acceleration. 

We examine with a particle-in-cell (PIC) simulation the formation of a 
shock in a plasma, in which a strong guiding magnetic field is quasi-parallel 
to the plasma expansion direction. Whistler waves, which become Alfv\'en 
waves at low frequencies, occuring at such shocks can be efficient electron 
accelerators \citep{Oblique3,Injection3,Gyrosurf,ANOTHERSLAM,Ken}. Whistlers 
and Alfv\'en waves are circulary polarized if they propagate parallel to the 
guiding magnetic field. We briefly summarize their properties and focus on the 
low-frequency modes with a left-hand polarization. These modes are 
qualitatively similar to the quasi-parallel propagating ones we consider
here. A more thorough description of the dispersion relation of two-fluid 
waves and the shift of the resonance frequencies for quasi-parallel 
propagation can be found elsewhere \citep{Treumann}. 

The dominant waves, which we will observe, grow in 
a plasma with an electron gyrofrequency that exceeds the plasma frequency. 
As we increase the frequency in such a strongly magnetized plasma towards
the ion cyclotron frequency, the Alfv\'en waves with a left-hand circular 
polarization become dispersive. These waves resonate with the ions and their 
frequencies remain below the ion cyclotron frequency. Alfv\'en modes with a 
left-hand circular polarization just below the ion cyclotron frequency are 
called ion whistlers. Whistlers are predominantly electromagnetic for small 
propagation angles relative to the guiding magnetic field, as it has been 
discussed for high-frequency ones by \citep{Tokar}, and if they have low 
wavenumbers. Ion whistlers or dispersive Alfv\'en waves develop a field-aligned 
electric field component close to the resonance frequency, by which they can 
interact nonlinearly with the plasma particles and accelerate them \citep{Ken}.
Any wave growth will furthermore result in an increasing energy density of 
the magnetic field.

Alfv\'en waves and other magnetohydrodynamic (MHD) waves can grow to amplitudes, 
at which they start to interact nonlinearly with the plasma \citep{STASI}. 
Short Large Amplitude Magnetic Field Structures (SLAMS) are nonlinear MHD 
waves occuring at quasi-parallel shocks and may be relevant for our 
simulation. The SLAMS can be efficient electron accelerators \citep{CME2}. 
Their magnetic amplitude reaches several times that of the background field 
and they can propagate with a super-Alfv\'enic speed, because they convect 
with the ions \citep{SLAMSPEED,SLAMS}. SLAMS have also been observed in 
simulations \citep{Scholer}. The acceleration of electrons by sub-structures 
of quasi-parallel shocks, which may be SLAMS, has been observed in the solar 
corona \citep{CME}. 

The absence of self-consistent kinetic models of oblique shocks implies 
that they can currently be studied only numerically with particle-in-cell 
(PIC) \citep{Code2,Code1} or with Vlasov simulations \citep{Arber,Sircombe}. 
The pioneering PIC simulations of plasma slabs, which collide with a speed 
of 0.9c and in the presence of an oblique magnetic field 
\citep{Oblique1,Oblique2}, have evidenced the formation of a shock that 
accelerated the electrons to ultrarelativistic speeds and amplified the 
magnetic field. A more recent PIC simulation study \citep{Shock1} has shown, 
that the shock formation is triggered by an energetic electromagnetic 
structure (EES). The simulation could demonstrate that an approximate 
equipartition of the ion, electron and magnetic energy densities is 
established. However, these simulations could only resolve one spatial 
dimension due to computer constraints, which is not necessarily realistic 
for mildly relativistic collision speeds.

We perform here a case study with initial conditions, which are similar
to those employed by \citet{Shock1}. We reduce the collision speed to 0.5c 
and lower the temperature. The plasma cloud representing the subshell is 
ten times denser than the plasma cloud that represents the ambient plasma 
(interstellar medium), into which the shell expands. A guiding magnetic 
field is quasi-parallel to the plasma flow velocity vector and it results 
in an electron gyrofrequency that equals the electron plasma frequency of 
the dense cloud. These bulk plasma parameters have been selected with the 
intention to enforce a planar shock front \citep{Nish,Mag,Shock1}, by which 
we can model the shock in one- or two-dimensions in space. The reduced 
ion-to-electron mass ratio we use allows us to model this collision in 
form of a 2(1/2)D particle-in-cell (PIC) simulation, which resolves two 
spatial and three momentum dimensions.

Our results are summarized as follows: We find higher-dimensional structures 
(density filaments), which initially form at the front of the tenuous plasma 
cloud and expand in time. The front of the dense cloud, which turns out to 
be the most relevant structure, remains planar and its filamentation is
delayed but not suppressed by the guiding magnetic field and the high 
temperature. An EES grows ahead of the dense plasma cloud before it has 
become filamentary and the magnetic amplitude of the EES reaches a value 
several times the one of the initial guiding magnetic field. The EES is 
pushed by the dense cloud and its phase speed in the rest frame of the 
tenuous cloud is comparable to the cloud collision speed or twice the 
Alfv\'en speed in the tenuous cloud. The front of the EES expands at an 
even higher speed. Its high speed and amplitude may imply that the EES is 
a SLAMS. The front of the dense cloud and, consequently, the EES are slowed 
down by the electromagnetic wave-particle interaction. It is thus not 
possible to define a rest frame moving with a constant speed, in which the 
EES is stationary. This would be necessary to measure the frequency of the 
EES accurately. However, the amplitude distribution of the EES suggests that 
its oscillation frequency is below the ion cyclotron frequency. 
Electromagnetic waves are destabilized by the EES ahead of the cloud overlap 
layer through what we think is a four-wave interaction \citep{Instability}.
The electrons are accelerated in the combined wave fields of these waves and
in the forming shock to highly relativistic speeds. The simulation shows
though that the strongest electron acceleration occurs at the position,
where the shock-reflected ion beam is forming. The plasma collision results 
in a substantial electron acceleration and also in an amplification of the 
magnetic energy density by one order of magnitude within the EES and the 
forming shock. Both values are probably limited by the reduced ion mass of 
400 electron masses, which we must employ. Radiative processes, which are 
not resolved by the PIC code, will at this stage start to influence the 
shock evolution \citep{Schl1,Schl2} and we stop the simulation.

This paper is structured as follows. Section 2 discusses the particle-in-cell
simulation method and the initial conditions. Section 3 presents our 
results, which are discussed in more detail in Section 4. 


\section{Initial conditions and simulation method}

\begin{figure}
\centering
\includegraphics[width=8cm]{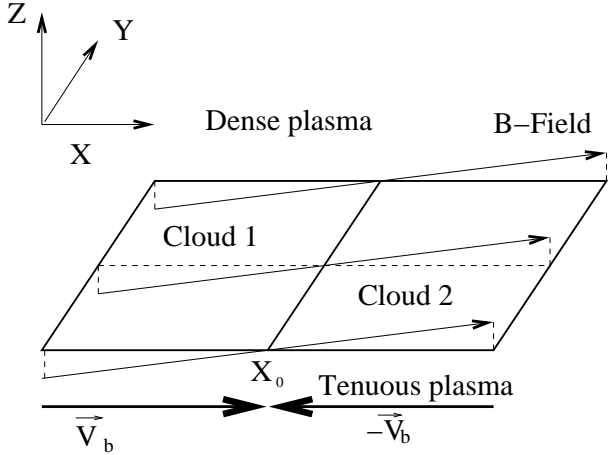}
\caption{Initial conditions. Two clouds are modelled. Each cloud occupies 
one half space and their mean velocity vectors $\vec{v}_{b}$ and $-\vec{v}_b$ 
point along the x-direction. The background magnetic field $\vec{B}_0 
\parallel (1,0,0.1)$. A convection electric field points along $\vec{y}$.}
\label{Fig1}
\end{figure}

We model the collision of two plasma clouds. Each cloud consists of 
co-moving and equally dense electron and ion species. This system is 
sketched in Fig.~\ref{Fig1}. 
The dense species represents in this model a fast subshell of the SNR 
blast shell and the tenuous species the ambient medium.

The number densities of the electrons with the charge $-e$ and mass $m_e$ 
of the dense cloud 1 and of the tenuous cloud 2 are $n_{e1}$ and $n_{e2} = 
n_{e1}/10$. The number densities of the ions with the charge and mass $e$ 
and $m_i$ of cloud 1 and 2 are $n_{i1}=n_{e1}$ and $n_{i2}=n_{e2}$. The 
cloud 1 occupies the half space $x< x_0$ and the cloud 2 occupies $x>x_0$.
Both clouds collide at the initial contact boundary, which we set to
$x_0 = 0$. All four species are initialized in their respective rest
frame with a relativistic Maxwellian distribution that has the temperature 
$T =$ 25 keV. This value exceeds by far the temperature of the ambient 
medium (ISM) into which the blast shell expands, which is typically less
than 1eV. The upstream temperature may, however, be increased by the
interaction of the ISM with the cosmic ray precursor of the main blast 
shell. The temperature $T$ is also higher than that downstream of the SNR 
shock, which is usually a few keV. The purpose of selecting this high
temperature of the simulation plasma is to improve the numerical efficiency. 
It maximizes the grid cell size, which must be comparable to the plasma 
Debye length, and the time step that is connected to the cell size, while 
it ensures that no highly relativistic particles are present to start with.

We set $v_b=0.268c$ and the cloud collision speed $v_c = 2v_b / (1+v_b^2/c^2)$ 
is $v_c = c/2$. The thermal speed $v_{e1} = {(k_B T / m_e)}^{1/2}$ and $v_{e2}
=v_{e1}$ of the electrons of the clouds 1 and 2 is $v_{e1} \approx 0.825 \, 
v_b$. The jump of the electron's mean velocity at $x=x_0$ is thus comparable 
to the thermal speed and the discontinuity in the electron phase space 
distribution is not strong. The thermal speeds $v_{i1} = v_{i2}$ of both
ion species are $v_{i1} = {(m_e / m_i)}^{0.5} v_{e1}$. The guiding 
magnetic field $\vec{B}_0 = B_{x,0} (1,0,0.1)$ with $eB_{x,0} / m_e = 
\Omega_{e1}$, where $\Omega_{e1} = {(n_{e1}e^2/m_e \epsilon_0)}^{1/2}$ is the 
electron plasma frequency of cloud 1. The convection electric field with 
the modulus $E_c = v_b B_{z,0}$ with $B_{z,0} = B_{x,0}/10$ points along 
$\vec{y}$ and changes its sign at $x=x_0$. The Alfv\'en speed in cloud 2 is 
$V_A \approx 0.23$ in units of $c$ and the collision has an initial Alfv\'en 
Mach number $\approx 2$. The $\beta = 2\mu_0 (n_{e1} + n_{i1}) k_BT / B^2_0 
= 0.2$, where $T$ is given in units of Kelvin, and the plasma is strongly 
magnetized. However, in the box frame of reference, the average kinetic 
energy density of the plasma exceeds the magnetic energy density by the 
factor 8.4 and the kinetic energy thus still dominates.

\begin{figure*}
\centering
\includegraphics[width=0.31\textwidth]{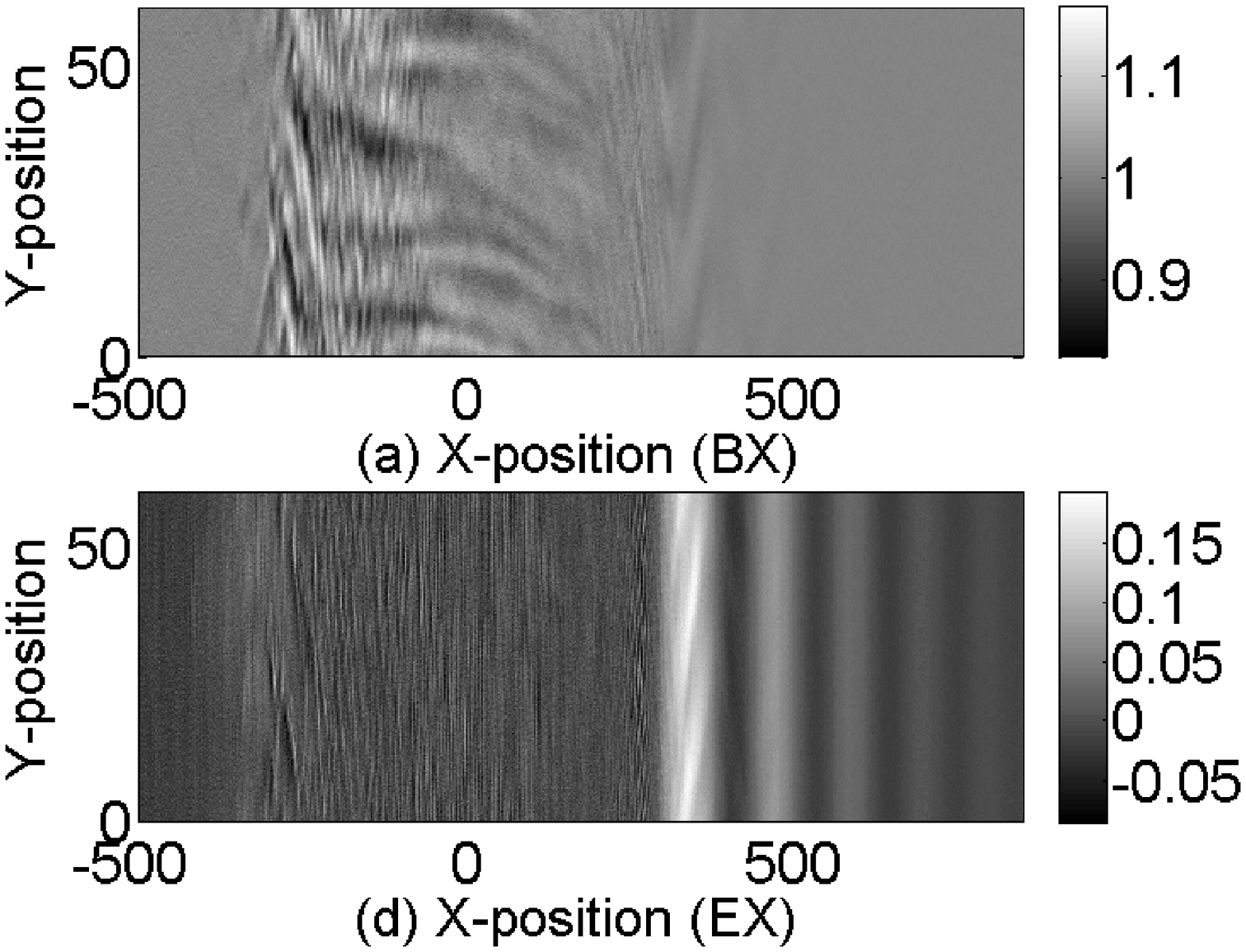}
\includegraphics[width=0.31\textwidth]{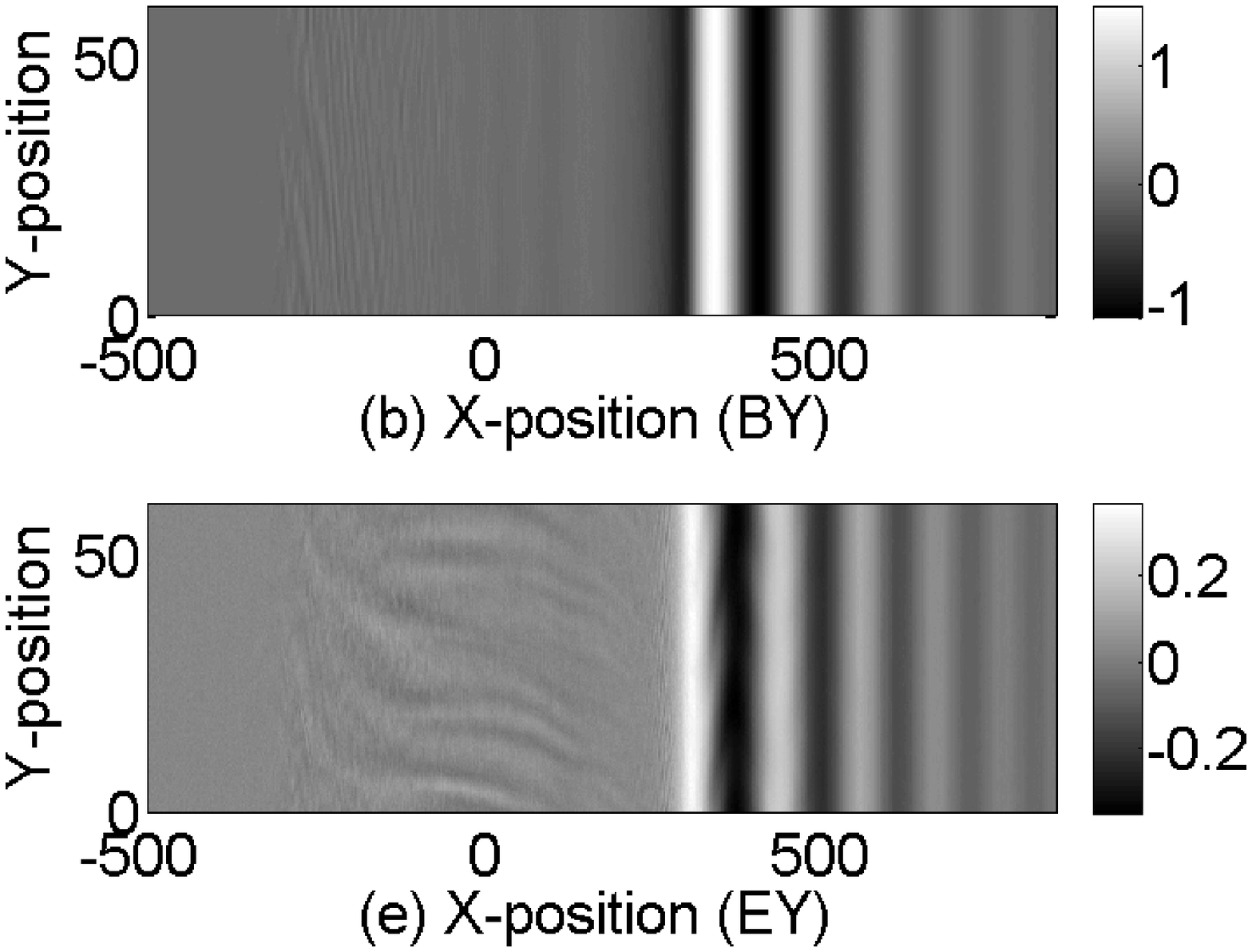}
\includegraphics[width=0.31\textwidth]{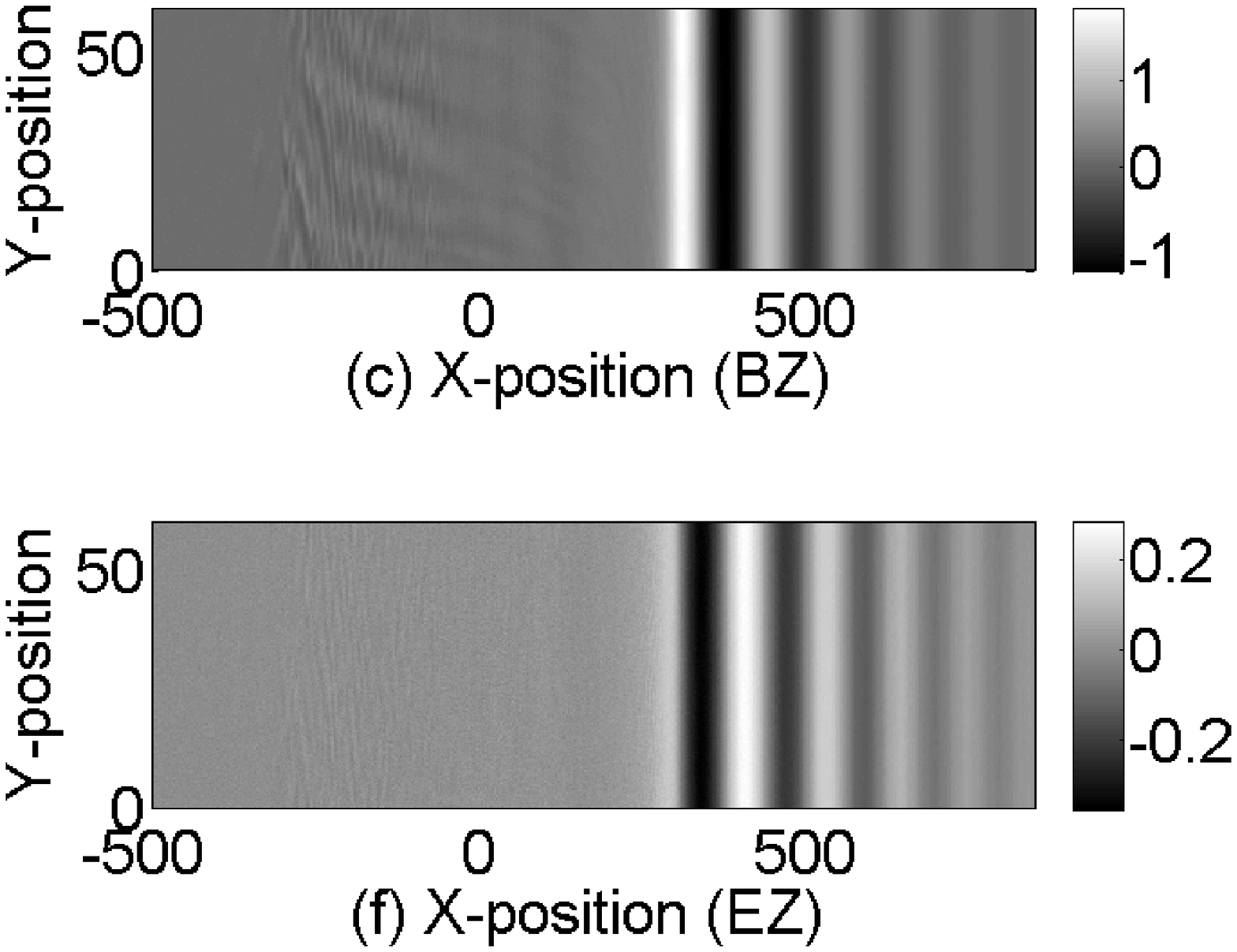}
\caption{Electromagnetic field amplitudes in the simulation at the time 
$T_1$. The upper row shows the magnetic field components $B_x (x,y)$ (a), 
$B_y (x,y)$ (b) and $B_z (x,y)$ (c). The lower row displays the $E_x (x,y)$ 
(d), $E_y (x,y)$ (e) and $E_z (x,y)$ (f).}\label{Fig2}
\end{figure*}

Typical particle number densities of the ambient interstellar medium, into 
which the plasma subshell would expand, are $0.1 \, cm^{-3} < n_0 < 1.0 \,
cm^{-3}$ \citep{Density,MagAmp1}. This ambient medium would correspond to 
the cloud 2. These densities are not always known, they change as a function 
of space and are not the same for all SNRs. We will thus use normalized 
units that permit us to scale the simulation results to any $n_0$. 

Variables in physical units are denoted by the subscript $p$. The time 
and space are scaled to $t = \Omega_{e1} t_p$ and $\vec{x} = \lambda_s^{-1} 
\vec{x}_p$, where $\lambda_s = c / \Omega_{e1}$ is the electron skin depth 
of cloud 1. The ion skin depth equals $20 \lambda_s$ for the considered
ion-to-electron mass ratio. The electric and magnetic fields are scaled to 
$\vec{E}_p (\lambda_s \vec{x}) = \Omega_{e1} c m_e e^{-1} \vec{E}(\vec{x})$ 
and $\vec{B}_p(\lambda_s \vec{x}) = \Omega_{e1} m_e e^{-1} \vec{B}(\vec{x})$. 
The charge, the particle number density and the current are normalized as 
$\rho_p (\lambda_s \vec{x}) = en_{e1} \rho (\vec{x})$, $n_p (\lambda_s \vec{x})
= n_{e1} n(\vec{x})$ and $\vec{J}_p (\lambda_s \vec{x}) = en_{e1}c 
\vec{J}(\vec{x})$. The normalized Maxwell's equations are 
\begin{eqnarray}
\nabla \times \vec{E} = -\frac{\partial \vec{B}}
{\partial t} & , & \nabla \cdot \vec{B} = 0, \\
\nabla \times \vec{B} = \frac{\partial \vec{E}}
{\partial t} + \vec{J} &,& \nabla \cdot \vec{E} = 
\rho.
\end{eqnarray}
A PIC code solves the Klimontovich-Dupree equations \citep{Dupree} that are 
derived from the Vlasov equation with the help of the method of characteristics.
The equations of motion 
\begin{eqnarray}
\frac{{\rm d}\vec{p}_j}{dt} = q_r \left ( \vec{E}(\vec{x}_j) + \vec{v}_j 
\times \vec{B}(\vec{x}_j) \right ), & \\ \vec{p}_j = m_r \gamma (\vec{v}_j) 
\vec{v}_j, \frac{{\rm d}\vec{x}_j}{dt} =\vec{v}_j
\end{eqnarray}
are solved for an ensemble of computational particles (CPs), where the 
subscript $j$ denotes one CP of the species $r$. The charge $q_r$ and 
mass $m_r$ of the CP are given in units of $e$ and $m_e$ and the momentum 
is normalized to $m_e c$. The macroscopic current $\vec{J}$ is obtained 
from interpolating the current contributions of each CP to the grid and 
from the summation of these interpolating currents over all particles. 
The macroscopic $\vec{J}$ is used to evolve $\vec{E}$ and $\vec{B}$ one
step in time. The momentum $\vec{p}_j$ of each CP with index $j$ is then 
updated with these new electromagnetic fields, which are interpolated to 
the position $\vec{x}_j$ of the respective CP. Interpolation schemes 
are discussed by \citet{Code2} and the one our code uses by \citet{Code1}.

Our code is relativistic and electromagnetic and we use periodic boundary
conditions in all directions. The simulation represents the $x,y$ plane. 
The simulation box length is $L_x = 5330$ along $x$, which is resolved by 
$1.8 \times 10^4$ grid cells. The length $L_y = 60$ along $y$ is resolved 
by 200 grid cells. Cloud 1 occupies the x-interval $-L_x/2 < x<0$, 
while cloud 2 occupies $0<x<L_x/2$. No new particles are introduced at the 
boundaries. The clouds convect away from the initial contact boundary at
$x=0$ with the speed modulus $v_b$ and the simulation is stopped before 
the rear boundaries of the clouds enter the area of interest.

Each plasma species is represented by 160 CPs per cell. The density 
difference of both clouds thus implies a lower weight for the CPs of cloud 2. 
We employ a mass ratio $m_i / m_e = 400$ and let the simulation run for the 
duration $T_{sim} = 3500$. If no strong instabilities develop, the particles 
could move on average a distance $v_b T_{sim}=950$.

\section{Simulation results}

In what follows, we analyse the data at two times. The field and particle
distribution is considered at the early time $T_1 = 1150$. We examine, 
which conditions eventually result in the growth of the shock. The shock 
is forming at the later time $T_{sim}$ and we determine the plasma
state that is reached by the release of the ion kinetic energy.

\subsection{Early simulation time} 

Figure~\ref{Fig2} displays the fields at the time $T_1$. The spatial 
interval $-300 < x < 300$, in which the clouds overlap by $v_b T_1 
\approx 300$, shows filamentary structures in $B_x$, $B_z$ and $E_y$. We 
discuss these first. The filamentation instability separates the currents 
due to both interpenetrating clouds and a current system develops in the 
simulation plane. It drives the $B_z$, which gives rise to the $E_y$ through 
linear \citep{Tzoufras} and nonlinear processes \citep{MagPre}. The $B_z$ 
is modulated along $y$ with a wavelength between $5 \lambda_s$ and $20 
\lambda_s$, the latter being equal to the ion skin depth. At least the 
largest structures of $B_z$ seem to be tied to ion filaments.  

Initially the convection electric field along $y$ ensures that the magnetic 
field is co-moving with the plasma. The gyrocenters of the plasma particles 
do not move relative to $\vec{B}_0$ and no net current is present. Once the 
plasma clouds overlap, there will be a net particle motion relative to 
$\vec{B}_0$ and the plasma particles can be deflected into the $y,z$-plane.
The derivative along $z$ vanishes in our 2D PIC simulation. The $B_x$ can 
only be generated through $\partial_x B_y - \partial_y B_x = \partial_t E_z 
+ J_z$, which causes structure formation in $B_x$ and $B_y$. Figure 
\ref{Fig2}(a) reveals that the filamentation of $B_x$ due to such currents 
is strongest close to the front of cloud 2, that is for $x<0$. The plots of 
$B_y$ and $E_z$ do not yet show a clear filamentation but one has to 
keep in mind that the apparent relative weakness of the modulations of $B_y$ 
compared to those of $B_x$ is influenced by the different colour scales. 
The fluctuations of $E_x$ in the interval $-300 < x < 300$ have no visible 
counterpart in $B_y$ and $B_z$. The polarization of the electric field
fluctuations along $\vec{v}_b$ suggests a charge density wave, but their 
correlation with the small-scale fluctuations of $B_x$ for $-300 < x < -100$ 
demonstrates that they are not purely electrostatic. 

The dominant fields of the energetic electromagnetic structure (EES) are 
observed in Fig. \ref{Fig2} ahead of the dense cloud 1 for $x>300$. The 
EES is planar. The field amplitudes along $y,z$ are practically constant 
along $y$. The $B_y$ and $B_z$ in Figs. \ref{Fig2}(b,c) reach peak 
amplitudes exceeding that of the guiding magnetic field in this interval. 
The $E_x$ and $B_x$ at $x\approx 300$ show an oblique modulation.

\begin{figure}
\centering
\includegraphics[width=8cm]{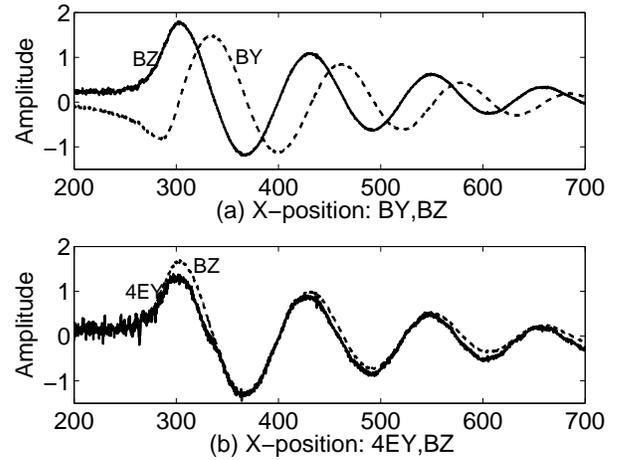}
\caption{Comparison of the planar electromagnetic fields along $x$ for 
$y=0$. (a) compares $cB_z$ with $cB_y$ in units of $cB_{x,0}$, while (b) 
compares $cB_z$ with $4E_y$ in the same normalization.}\label{Fig3}
\end{figure}

We exploit the planarity and constancy as a function of $y$ of the 
amplitudes of $B_y,B_z$ and $E_y$ for $x>250$, where we find the EES. 
Figure~\ref{Fig3} plots these amplitudes as a function of $x$ for $y=0$.
The considered field components have their peak values at the front 
of cloud 1 located at $x \approx 300$. The amplitudes decrease
exponentially for $x<300$ and $B_y$ and $B_z$ reveal a phase shift of
$180^\circ$. The amplitudes of $B_y$, $B_z$ and $E_y$ decrease also for 
increasing $x>300$, although more slowly and the fields oscillate. The 
phases of $B_y$ and $B_z$ are shifted by $\approx 90^\circ$ in this 
interval and the EES is predominantly 
circularly polarized. The electron gyrofrequency is comparable to the plasma 
frequency for $x<300$ and significantly larger for $x>300$. The jump in the 
ratio of the characteristic resonance frequencies at $x\approx 300$ will 
alter the wave dispersive properties of the plasma, explaining the qualitative
change the fields of the EES undergo as we cross this boundary.

The $B_y$ and $B_z$ of the EES have a left-hand polarization for $x>300$. 
Consider a coordinate system defined by $x$, $B_y$ and $B_z$. The $B_y$ 
has a maximum at $x\approx 340$, where $B_z = 0$. As we go to increasing 
values of $x$, the amplitude of $B_y$ decreases and that of $B_z$ becomes 
negative. The magnetic field vector rotates counter-clockwise. A deviation 
from a circular polarization is provided by $B_{z,0} \neq 0$ and by $E_x 
\neq 0$, but their amplitudes are relatively low compared to $B_y$ and $B_z$ 
for $300 < x < 600$. 

The $B_z$ is almost in phase with $E_y$. The comparison of $B_y$ with $E_y$ 
and the comparison of $B_z$ with $E_z$ for $x> 300$ would show that they 
are shifted by $\approx 90^\circ$, while $B_y$ is shifted by $180^\circ$ 
relative to $E_z$. Figure \ref{Fig2} shows this most clearly at $x\approx 
500$. The wave components $\vec{E}_{\perp} = (0,E_y,E_z)$ and $\vec{B}_{\perp} 
= (0,B_y,B_z)$ perpendicular to $\vec{v}_b$ can thus be connected through 
$\vec{E}_{\perp} \approx -\vec{v}_b \times \vec{B}_{\perp}$, where we assume
that $\vec{v}_b \parallel \vec{x}$. The EES and its magnetic field is thus 
convecting with the plasma of cloud 1. The wavelength of $E_x$ in Fig. 
\ref{Fig2} along $x$ equals that of all field components except $B_x$ for 
$x>300$. Since all components of $\vec{E}$ are strong, the EES has an 
electric field component aligned with $\vec{B}_0$. The EES has the wavelength 
$\lambda_W \approx 120 \lambda_s$, coinciding with $\lambda_W \approx 2\pi 
{(m_i / m_e)}^{1/2} \lambda_s$. This $\lambda_W$, together with a convection 
of the EES with the $v_b$ of cloud 1, points at a connection of the EES to 
phase space structures in the ion distribution, which is typical for SLAMS 
\citep{SLAMS}.

The magnetic oscillations observed in Fig. \ref{Fig2} in the spatial 
interval $-300 < x < 300$, in which both clouds overlap, must be tied 
to currents. The currents originate from a spatial redistribution of 
the plasma cloud particles and the latter can be examined with the help 
of the density distributions of the four plasma populations. 

\begin{figure}
\centering
\includegraphics[width=\columnwidth]{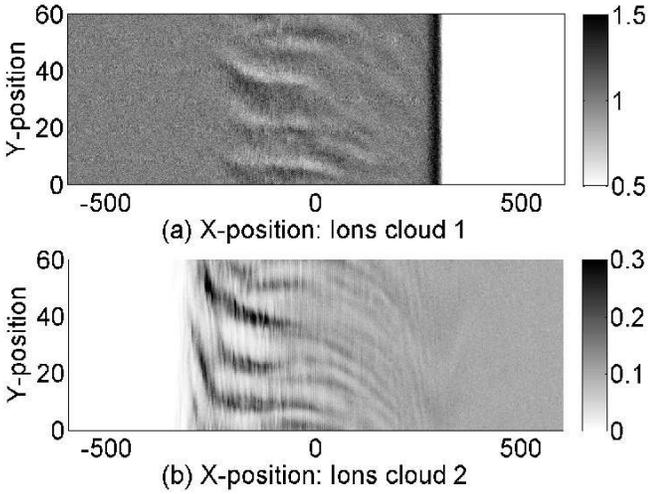}
\caption{Ion densities at the time $T_1$ normalized to $n_{i1}$. The ions of 
cloud 1 are shown in (a) and those of cloud 2 in (b).} 
\label{Fig4}
\end{figure}

Figure~\ref{Fig4} displays the densities of the ions of cloud 1 and of 
cloud 2.
The ion density of cloud 1 has a peak at $x \approx 300$. This high-density 
structure is planar and shows no signs of filamentation along $y$. The 
filamented front of cloud 2 at $x \approx -300$ shows no high-density 
structure. The ion densities of both clouds are only strongly modulated 
well behind the expansion front of cloud 1. Modulations of the density 
of cloud 1 of up to 40\% are observed for $-200 < x < 0$ and the density 
of cloud 2 is modulated even stronger for $-300<x<0$. Weaker modulations of 
both ion clouds extend up to $x\approx 200$ and their wavelength decreases 
with increasing $x$. The filaments of both ion clouds are aligned 
with the $x$-axis in the interval $-250 < x < -100$. The ion density of 
cloud 1 is highest at $y\approx 0$ and lowest at $y\approx 10$, while the 
density of cloud 2 shows the opposite behaviour. The ions of both clouds 
and their currents are spatially separated, as expected. The ion 
filamentation takes place on the same scale as the oscillations of $B_z$ in 
Fig. \ref{Fig2}(c) in the same interval, reinforcing their connection. The 
ion density modulation of cloud 2 oblique to the $x$-axis in the interval 
$-300 < x < -250$ is not balanced by the ions of cloud 1.

It is remarkable that the amplitude of $B_z$ in the interval $-300 < x < 0$, 
which displays the strongest ion filamentation, is an order of magnitude 
lower than that of the EES. The latter must thus have a different origin.

\begin{figure}
\centering
\includegraphics[width=\columnwidth]{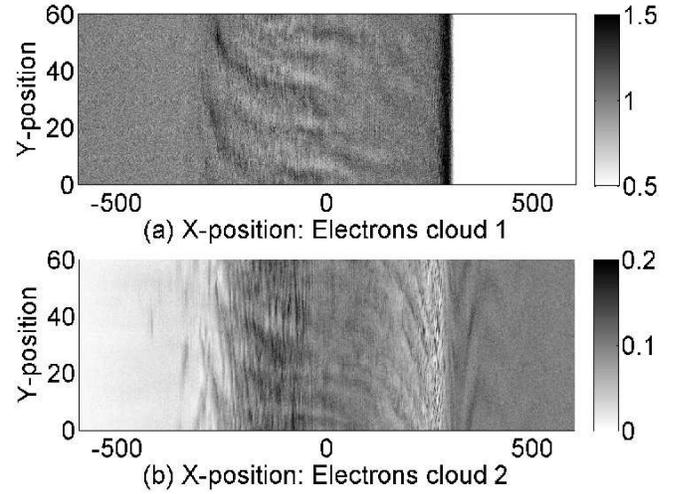}
\caption{Electron densities at the time $T_1$ normalized to $n_{e1}$. The  
electrons of cloud 1 are shown in (a) and those of cloud 2 in (b).}
\label{Fig5}
\end{figure}

Figure~\ref{Fig5} shows the density distributions of the electrons of 
cloud 1 and 2.
The electrons of cloud 1 accumulate at $x\approx 300$ and their density
is modulated on scales of an ion skin depth in the interval $-300 < x < 0$. 
Similar structures are revealed by the electron density of cloud 2 in this 
interval. These patterns follow closely those of the ions in Fig.~\ref{Fig4}, 
which is at least partially enforced by the need to satisfy the 
quasi-neutrality of the plasma. This is evidenced, for example, by the 
electrons of cloud 1 for $-300 < x < -250$. Their density shows the same 
oblique striping as the ion density of cloud 2 in Fig. \ref{Fig4}(b). The 
density maxima of the ions of cloud 2 and of the electrons of cloud 1, as 
well as their minima, follow each other closely in this interval. It is the 
electrons, rather than the ions of cloud 1, which compensate the net 
ion charge in this interval. Another consequence of the quasi-neutrality 
is seen at the front of cloud 1, where the electrons reach a high density. 
This density peak is well-defined and confined to a narrow $x$-interval, 
despite the high thermal electron speed, which should rapidly disperse them. 
The high ion density in this interval provides the confinement. In contrast, 
the cloud 2 shows no ion density enhancement in Fig. \ref{Fig4}(b) at its 
front at $x\approx -300$. Consequently, the electrons of cloud 2 are spread 
out along $x$ and the fastest electrons have reached an $x < -500$. 

The ions and electrons of cloud 2 in Fig.~\ref{Fig4}(b) and 
Fig.~\ref{Fig5}(b) show patterns in their density ahead of cloud 1 for 
$x>300$. These patterns resemble those in $B_x$ and $E_x$ at $x \approx 
300$ in Fig.~\ref{Fig2}. The box width along $y$ upstream of cloud 1 is 
approximately one ion skin depth of cloud 2, because $n_{i1}/n_{i2} = 10$. 
The structures just ahead of $x=300$ perform two oscillations along $y$. 
These are potentially filaments, but they cannot be caused by the 
interaction of cloud 1 and 2 because they are outside the cloud overlap 
layer. The spatial modulation provides a hint. The filament starting at 
$x \approx 300$ and $y\approx 30$ in Fig. \ref{Fig5}(b) changes the value 
of $y$, at which the density is highest, continuously as a function of $x$. 
One oscillation takes place on a distance $\approx 120$ along $x$, which is 
the wavelength $\lambda_W$ of the EES.

\begin{figure}
\centering
\includegraphics[width=\columnwidth]{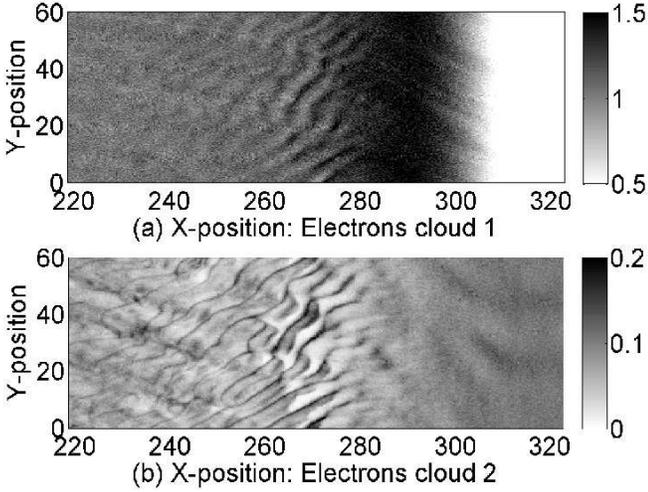}
\caption{Electron densities at the time $T_1$ close to the front of cloud
1 and normalized to $n_{e1}$. The electrons of cloud 1 are shown in (a) and 
those of cloud 2 in (b).}\label{Fig6}
\end{figure}

The electrons of cloud 2 in Fig. \ref{Fig5}(b) close to the front of the 
cloud 1 at $x\approx 300$ are also modulated on a smaller scale, which
is already evident from Fig. \ref{Fig5}. Figure \ref{Fig6} concentrates
on this interval. 
Filaments can be seen with a thickness ranging from $\lambda_s$ to about 
$5\lambda_s$ close to $x\approx 280$ and they have vanished already 
a few tens of $\lambda_s$ behind the front of cloud 1. Their size and 
even more so their short lifetime due to a rapid thermalization indicates 
that these structures are caused by a filamentation of the electrons. 
Figure~\ref{Fig4}(b) confirms this, because the large ion filaments form 
far behind the front of cloud 1, implying that the ions have not been 
thermalized by their passage through the interval depicted by Fig.~\ref{Fig6}.

\begin{figure}
\centering
\includegraphics[width=\columnwidth]{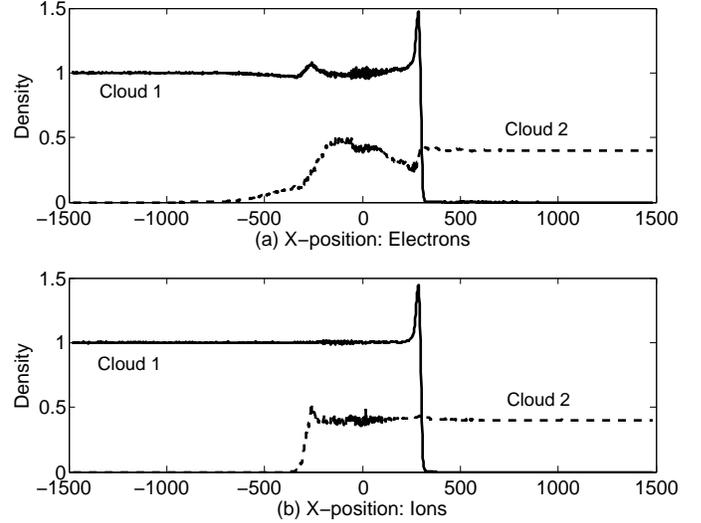}
\caption{Densities at the time $T_1$ integrated over $y$. (a) shows the 
electron density and (b) that of the ions. The dashed curves correspond 
to the electrons and ions of the cloud 2 and their densities are multiplied 
by the factor 4.}
\label{Fig7}
\end{figure}

The densities of the four species in Fig.~\ref{Fig7} are integrated over $y$.
The plasma density of cloud 1 reaches 1.5 times its initial value at its 
front at $x\approx 300$ and it rapidly vanishes for further increasing $x$. 
No ions of cloud 1 and almost none of its electrons have thus propagated 
upstream of its front. The plasma density of cloud 2 is almost constant for 
$x>300$. The electron gyroradius in $\vec{B}_{\perp}$ is less than $\lambda_s$ 
and they should fall behind the ions of cloud 2 once they encounter the EES. 
One may expect that the electrons of cloud 2 accumulate at this position, 
this is however not seen in Fig. \ref{Fig7}. The absence of electrons of 
cloud 1 at this position and the need to satisfy the plasma quasi-neutrality 
transports the electrons of cloud 2 across $\vec{B}_{\perp}$, which will 
accelerate these electrons orthogonally to $x$.

The ion density at the front of cloud 2 and the electron density of cloud
1 at $x\approx -300$ are weakly enhanced. These increased densities are
caused by the oblique structures in the interval $-300 < x < -250$ in 
Fig.~\ref{Fig4}(b) and Fig.~\ref{Fig5}(a). The velocity vectors of the 
ions moving in the oblique filaments have been deflected away from 
$\vec{v}_b$. This decreases their flow speed along $x$ and the ions 
accumulate.

The electron densities of both clouds are nonuniform within $-600 < x < 300$,
evidencing their interactions through instabilities. The density dip of
the electrons of cloud 2 for $250 < x < 300$ coincides with the filamentary 
structures in the same interval in Fig.~\ref{Fig6}, further evidencing that 
this dip is caused by an electron filamentation instability without a
significant involvement of the ions. A notable fraction of the electrons of 
cloud 2 have 
convected to $x\approx -600$ and some even beyond that. The densities of 
both ion clouds do not yet show a significant modulation, apart from the 
accumulation at $x\approx 300$. The ions in the layer within $-300 < x < 0$ 
in Fig. \ref{Fig4} are only redistributed in the $x,y$-plane, but they do 
not accumulate along $x$ through a shock compression. 

The interaction of the particles with the electromagnetic fields implies
that their dynamics involves at least one spatial and three momentum
dimensions. We display the phase space projections, which we consider to
be the most relevant ones. The ion phase space density $f_i (x,p_x)$ can be 
used to determine if and when the shock forms. The ions of cloud 1 and 2 
must merge to form the downstream region. The electron distribution  
$f_e (x,\gamma)$ shows the range of Lorentz factors attained by the electrons.

\begin{figure}
\centering
\includegraphics[width=\columnwidth]{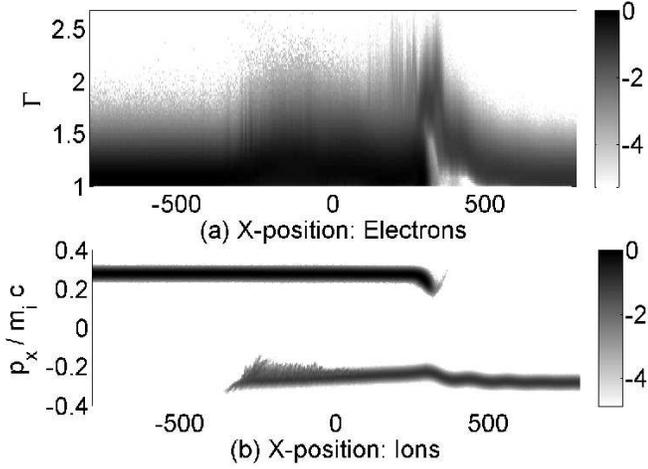}
\caption{Phase space densities at the time $T_1$. (a) corresponds to electrons
and (b) to ions. The greyscale denotes the base 10 logarithm of the density.}
\label{Fig8}
\end{figure}

Figure~\ref{Fig8} displays these distributions.
The ions of cloud 1 convect practically freely. Only the tip of this cloud 
at $x\approx 300$ is modified. The electron gyrofrequency $\omega_{ce} = 
e|\vec{B}_0|/ m_e$ equals $\Omega_{e1}$, with which we have normalized the 
time. The ions have gyrated $T_1 / 400 \approx \pi$ or one half of a 
gyro-orbit in $\vec{B}_0$ until now. The uniform $\vec{B}_0$ is, however, 
almost parallel to $x$ and it cannot cause by itself the ion rotation at the 
front of cloud 1. The magnetic field $B_\perp = {(B_y^2 + B_z^2)}^{1/2}$ of 
the EES exceeds $B_{x,0}$ at the front of cloud 1. The ions will have gyrated 
by a significant angle in this $\vec{B}_\perp$, because the amplitude of the 
EES grows rapidly to a large amplitude. The EES also accelerates the 
electrons of cloud 2 in the interval $300 < x < 500$, which are transported 
across $\vec{B}_\perp$ by the ions of cloud 2. The ions of both clouds must 
provide through the EES the energy required to accelerate the electrons. The 
modification of the ion phase space distribution can thus be attributed to 
the EES and to the electron acceleration.

Weak oscillations of the mean momentum $\langle p_x \rangle (x)$ can be seen 
in the ions of cloud 2 in the interval $300 < x < 600$ in Fig.~\ref{Fig8}. 
We can interpret them as follows. The ions of cloud 2 upstream of cloud 1
move through the EES at the speed $\approx -v_b$ in the box frame towards 
decreasing values of $x$. A strong circularly polarized magnetowave with a 
wavevector aligned with $x$ would force the ions onto a circular path in 
the $p_y,p_z$ plane, in which the mean ion beam momenta $\langle p_y \rangle 
(x)$ and $\langle p_z \rangle (x)$ vary as a function of $x$ with the wave's
periodicity. Since $B_{z,0}$ is not negligible, this motion must also modulate 
$\langle p_x \rangle (x)$. Finally we see that the ions of the tenuous cloud 2 
have been heated up by their filamentation at the front of cloud 2 at 
$x \approx -300$. 

\subsection{Late simulation time}

Unless stated otherwise, we analyse now the simulation data at the time 
$T_{sim}=3500$, when we stop the simulation. If both beams had streamed 
freely, they should have propagated on average the distance $v_b T_{sim} 
\approx 950$.
 
\begin{figure}
\centering
\includegraphics[width=\columnwidth]{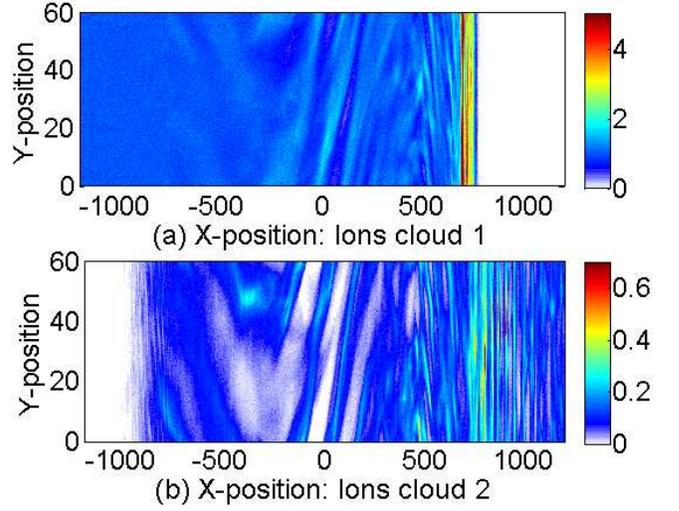}
\caption{(Colour online) Ion density distributions at the time $T_{sim}$ 
in units of $n_{i1}$. The ions of cloud 1 are shown in (a) and the ions 
of cloud 2 in (b).}\label{Fig9}
\end{figure}

Figure~\ref{Fig9} displays the densities of both ion clouds.
The front of the ions of cloud 1 is located at $x \approx 750$, which is
well behind $v_b T_{sim} \approx 950$. The ions of cloud 1 have thus been
slowed down and the peak density reaches a value $\approx 6$ at $x\approx 750$. 
This high-density structure is still planar, but its density is not 
constant as a function of $y$ any more. It decreases to $4$ at 
$x \approx 750$ and at $y \approx 40$. The front of the ions of 
the cloud 2 is located at $x \approx -950$, which equals the distance 
expected from the free streaming of the ions. Both ion clouds reveal 
density modulations on a scale less or comparable to the ion skin depth. 
The largest is found in the ions of cloud 2 at $x\approx -300$ and the 
interval with a reduced density ranges from $y \approx 0$ to $y\approx 30$. 
The ions of cloud 2 also show structures in the density distribution for 
$x > 750$, which is outside the cloud overlap layer. The wavevector of 
these oscillations is parallel to $x$.

Movie 1 shows the time evolution of the ion densities until the time $T_{sim}$.
We first discuss the filamentation in the cloud overlap layer. The filament 
formation is seen clearly, when each cloud has expanded by the distance 100 
or for a time $T_1/3$. The filaments are asymmetric, a consequence 
of the unequal cloud densities. The ions of cloud 2 show at this time and 
in the interval $-100 < x < 0$ several filaments aligned with the $x$-axis. 
The filaments have a constant density along $y$ and a width $\approx 10$. 
These filaments are separated by intervals with a reduced density, which 
have a width of only a few $\lambda_s$. The latter have counterparts in
the ions of cloud 1. The wave vector of these filaments is parallel to the 
$y$-axis at this time.

We also observe density fluctuations of the ions of cloud 2 with a wave
vector along $x$, which develop first in the section of the cloud overlap 
layer with $x<0$. This is an instability developing as the ions of cloud 2
propagate through the plasma of cloud 1. A Buneman-type instability 
\citep{Buneman} develops between the ions of cloud 2, which can be considered 
to form an unmagnetized beam on electron time scales, and the magnetized
electrons of cloud 1. The phase space density oscillations of the ions of 
cloud 2 are periodic with a wave length $\lambda \la 4 \pi v_b / \Omega_{e1}$ 
(not shown), which is typical for the Buneman instability involving hot 
electrons. The 
phase speed of the unstable wave is close to the speed of the ion beam, here 
composed of the ions of cloud 2, turning it into a slow oscillation in its 
rest frame. The ions can react to it and they will form phase space holes 
after a sufficiently long time. The ions of cloud 1 see the Buneman wave as 
a rapid oscillation to which they cannot react. This explains why the ions 
of cloud 2 in movie 1 at the time $T_1$ and the electrons in Fig.~\ref{Fig5} 
show the density fluctuations, but not the ions of cloud 1 in movie 1. These 
charge density fluctuations give the oscillations of $E_x$ along $x$ in 
Fig.~\ref{Fig2}(d). The electrostatic fields probably couple to $B_x$, 
because the Buneman instability is not purely electrostatic in the presence
of the strong oblique $\vec{B}_0$.
  
The ion filaments are observed to move in time in the positive $y$-direction, 
when the clouds have expanded by the distance 100 along $x$. The flow 
pattern becomes complicated, when cloud 1 has expanded to $x\approx 400$. 
The filaments in the interval $0 < x < 300$ still convect to increasing $y$, 
while the ion filaments of cloud 2 with $x<0$ move in the opposite direction. 
We may interpret this observation as follows. Only the ions in the cloud 
overlap layer are not co-moving with $\vec{B}_0$, because we cannot define 
through the convection electric field a reference frame, in which the ions 
of both clouds are at rest. The ions in the cloud overlap layer rotate with 
a velocity amplitude set by their mean speed relative to the rest frame of 
the overlap layer, rather than by their smaller thermal speed outside it. 

The ions gyrate in the cloud overlap layer orthogonally to $\vec{B}_0$ in a 
plane that is almost parallel to the $p_y,p_z$ plane. The ions of cloud 1 
and cloud 2 are counter-propagating and the magnetic field should deflect 
them into opposite directions. The $B_z$ separates the filaments of both 
ion clouds. This implies in a 2D geometry that the filaments are locked as 
long as the magnetic trapping force by $B_z$ \citep{Davidson} is stronger 
than the Lorentz force due to $\vec{B}_0$. A common direction of motion 
is established. The direction of motion that varies with $x$ and later also 
with $y$ is probably tied to the ion gyro-phase. Movie 1 suggests at late 
times that the filaments of cloud 2 close to its front merge and separate 
again. The $B_z$ is here insufficiently strong to separate the ion filaments
in the $x,y$ plane. The $\vec{B}_0 \neq 0$ introduces a complicated flow 
pattern in the 5D phase space spanned by $x,y$ and by the three components 
of $\vec{p}$. The filaments do probably not merge in this high-dimensional 
space but they appear to do so in the simulation plane. 

Now we turn towards the filaments upstream of cloud 1. As we approach the
time $T_1$, when the clouds have expanded by the distance 300 along $x$ in 
the movie 1, the filaments in the cloud overlap region have fully developed 
and we observe first signs of a filamentation of the cloud 2 ahead of the 
front of cloud 1. These filaments are initially oblique as Fig.~\ref{Fig4}(b) 
depicts. When cloud 1 has reached $x\approx 600$, the upstream region ahead 
of cloud 1 shows strong density oscillations of the ions of cloud 2 along 
the $x$-axis, which are also depicted in Fig.~\ref{Fig9}(b). They cannot be 
related directly to the EES, because their wavelength along $x$ is about 
$\lambda_W/3$. The patterns and their time-evolution eventually become 
complicated and presumably also high-dimensional.

\begin{figure}
\centering
\includegraphics[width=\columnwidth]{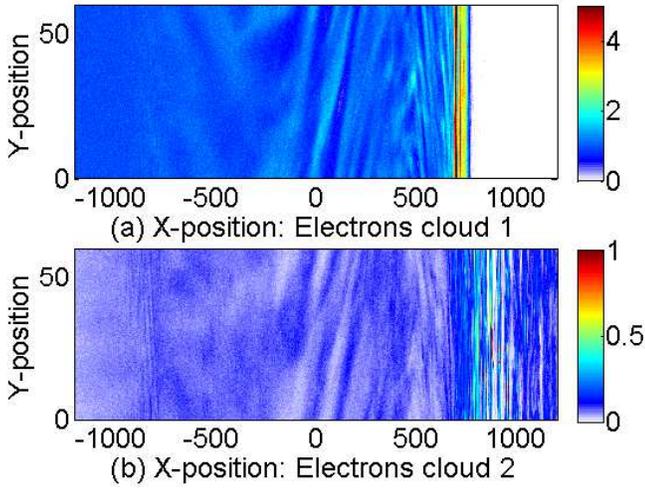}
\caption{(Colour online) Electron density distributions at the time $T_{sim}$
in units of $n_{e1}$. (a) displays the electrons of cloud 1 and (b) the
electrons of cloud 2.}
\label{Fig10}
\end{figure}

The density distributions of the electrons of both clouds are correlated 
with those of the ions in the overlap region $-950 < x < 750$, as the 
Fig.~\ref{Fig10} is evidencing. An example is here the density pattern, 
which crosses $y=0$ at $|x| < 100$ and is visible in all four species. 
The filaments in the interval $-500<x<500$ are, however, more diffuse 
than those of the ions in Fig.~\ref{Fig9}. The density peak of the 
electrons of cloud 1 at $x\approx 750$ is comparable to that of the ions. 
Again, the electrons are confined to maintain the quasi-neutrality of 
the plasma. The front end of the electrons of cloud 2 has been 
replaced by a gradual decrease of the electron density between $-1500 
< x < -800$. The electrons with $x<-950$ are the fastest ones of cloud 2, 
which have outrun the bulk of the electrons moving with $-v_b$. The 
ripples in the electron density of cloud 2 at $x>750$ are practically 
identical to those of the ions in Fig.~\ref{Fig9}(b).

\begin{figure*}
\centering
\includegraphics[width=0.31\textwidth]{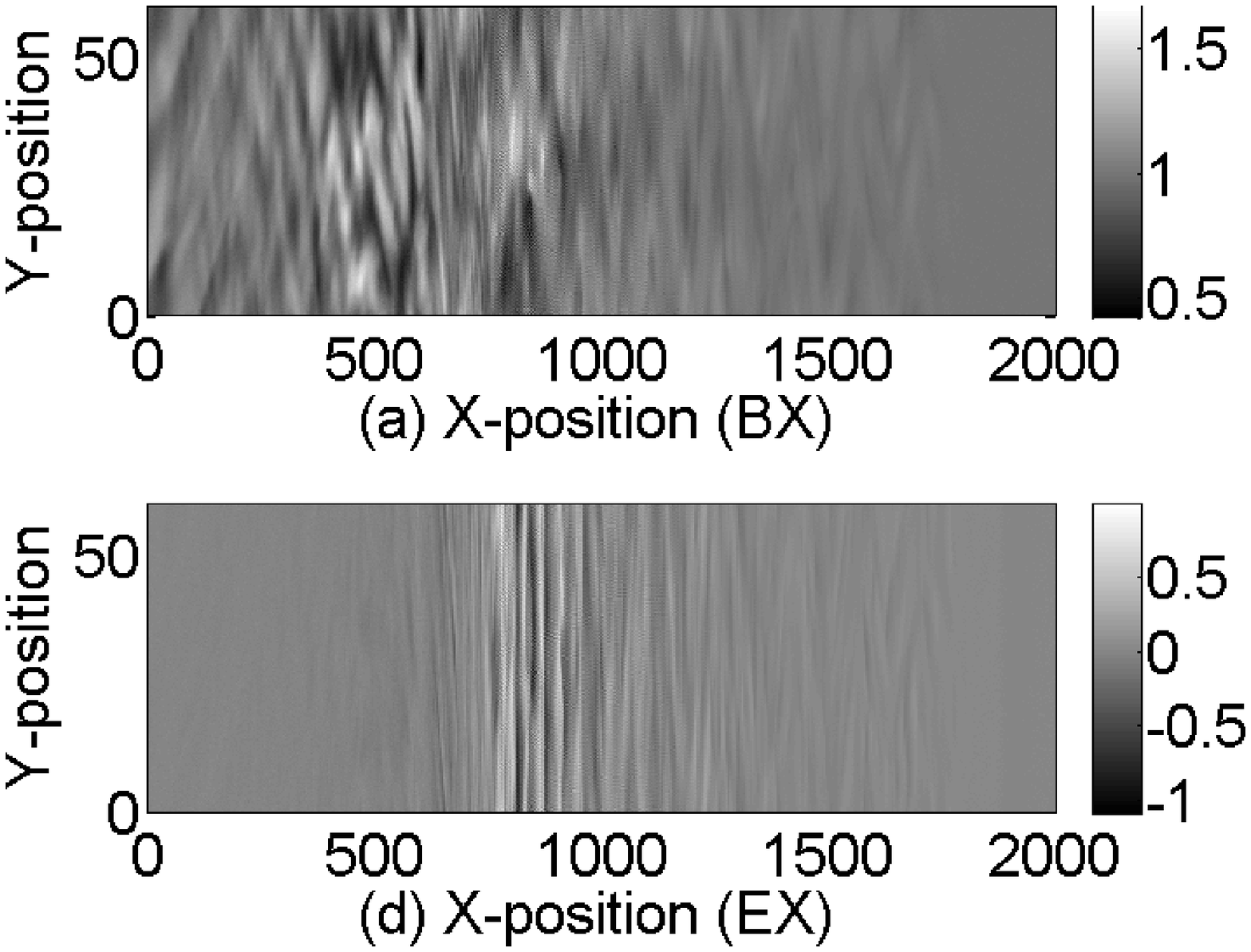}
\includegraphics[width=0.31\textwidth]{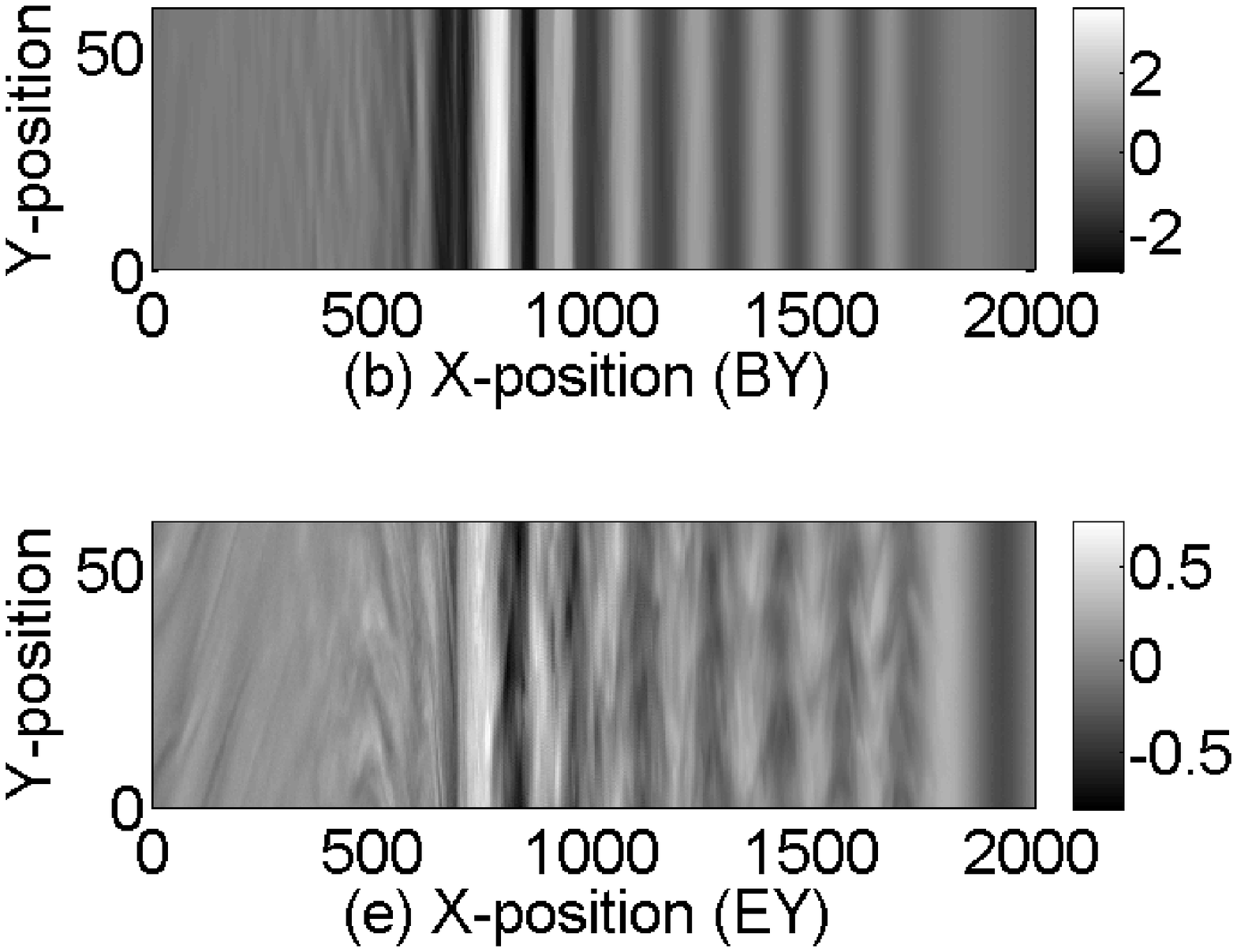}
\includegraphics[width=0.31\textwidth]{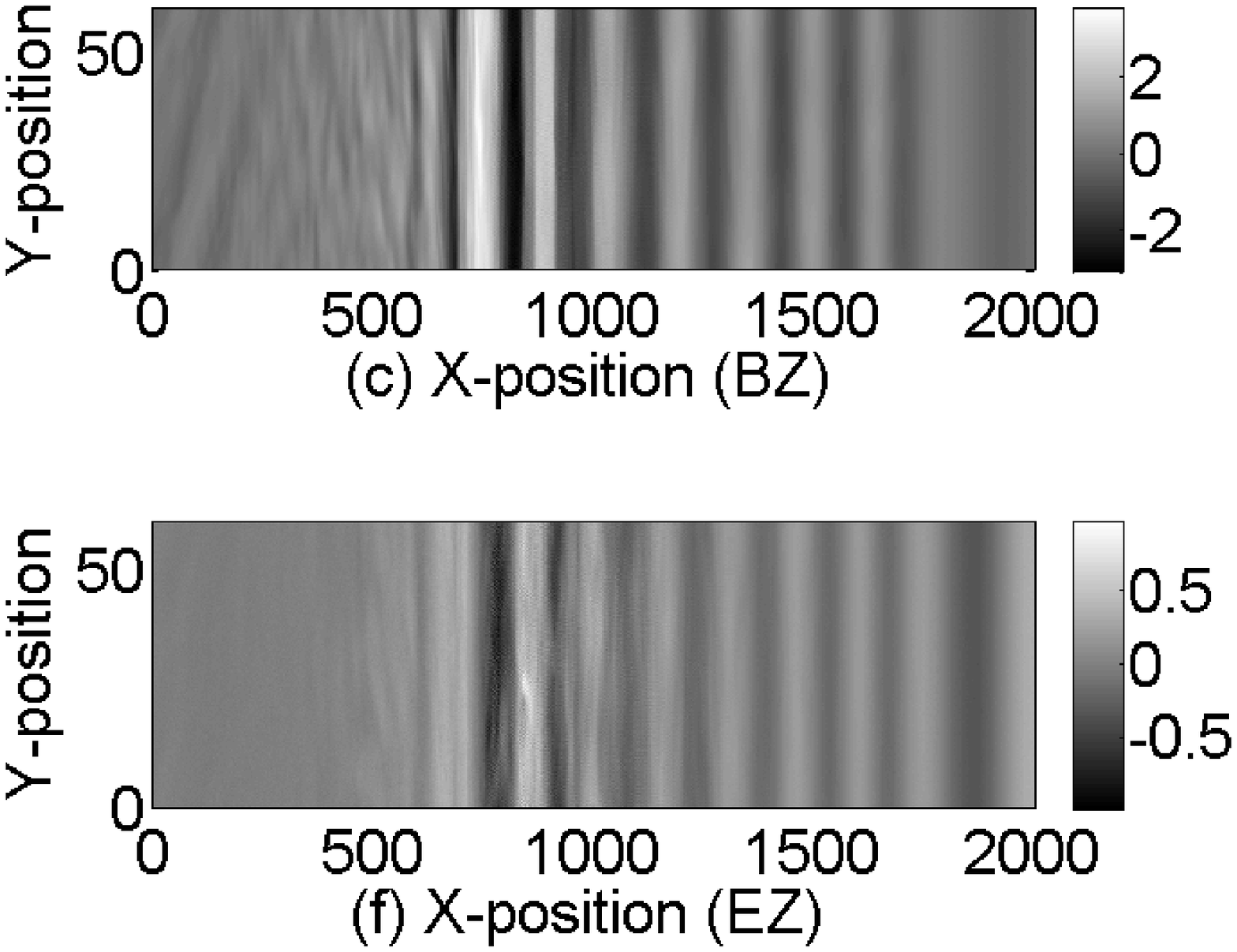}
\caption{Electromagnetic field amplitudes at the time $T_{sim}$. The upper 
row shows the magnetic field components $B_x$ (a), $B_y$ (b) and $B_z$ (c). 
The lower row displays the $E_x$ (d), $E_y$ (e) and $E_z$ (f).}\label{Fig11}
\end{figure*}

Movie 2 reveals the time-evolution of the electron densities. Density 
structures form rapidly close to the initial 
cloud contact boundary. These structures have no counterpart in the ion 
distributions in movie 1, until the cloud overlap layer covers $|x| \la 50$. 
This initial plasma evolution is thus determined by the electron 
filamentation. The density of the electrons of cloud 1 is enhanced in 
the cloud overlap layer, while the electron density of cloud 2 is decreased
(See also Fig.~\ref{Fig7}(a)). Already at this early time, the fastest 
electrons of cloud 2 outrun the bulk of the electrons to form a low-density 
tail along $x$ and one consequence is the reduction of the electron density 
of cloud 2 in the cloud overlap layer. A strong correlation between the ions 
of both clouds and the electrons of cloud 1 in the overlap layer is observed, 
when the latter covers $|x| \la 100$, demonstrating that these three species 
are involved in the formation of the large scale pattern. The electrons of 
cloud 2 show a distinct behaviour. This is also observed at a later time, 
when the overlap layer covers $|x| \la 200$. The electrons of cloud 2 are 
redistributed at this time immediately after they have crossed the front 
of cloud 1 at $x\approx 200$, which is also shown by Fig.~\ref{Fig6}, and 
their density is not visibly correlated with those of the other species. 

This behaviour can be understood qualitatively. Consider a dense electron 
beam and a tenuous electron beam that undergo the filamentation instability, 
which separates their currents. A highly nonlinear filamentary electron 
distribution of the tenuous beam can be compensated by a weak perturbation 
of the dense beam. We thus expect that the spatial density modulations of 
the tenuous beam become nonlinear first. We also expect, that the high ion 
inertia results in a delay of their filamentation compared to that of the 
electrons and both species of cloud 2 will interact differently with the 
plasma of cloud 1. 

The ions of cloud 2 in Movie 1 show structures similar to those in cloud 1, 
which have initially no counterpart in the electrons of cloud 2 (Movie 2). 
The ions of cloud 2 reveal a higher degree of non-uniformity than the 
electrons of cloud 1, when the overlap layer spans the interval $|x| \la 
200$. We can attribute this to their lower mean density. The increasingly 
pronounced large-scale modulation of the plasma density of cloud 1 at the 
time $T_1$ imprints itself onto the electrons of cloud 2. The latter show 
a modulation on two scales in Fig. \ref{Fig5}. The density evolution of the 
electrons and of the ions eventually becomes similar, except close to the 
front of cloud 1. Far behind this front, the electrons have thermalized. 
Since they have no free energy left which could lead to independent behaviour 
their density distribution follows that of the ions.

We explore now the density evolution of the electrons and ions of cloud 2 
ahead of the front of cloud 1 in the movies 1 and 2. We determine now, if 
we find multiple particle beams upstream of cloud 1 that could give rise
to beam-driven instabilities. The electrons of cloud 2 
thermalize after they have crossed the front of cloud 1 in Fig. \ref{Fig6} 
and we can no longer distinguish between the electrons of either cloud. 
Movie 2 shows no electrons of cloud 1 leaking upstream and the same should 
hold for the electrons of cloud 2, which have been mixed with those of cloud 1. 
This is supported by Fig.~\ref{Fig7}, which demonstrates firstly that no 
significant numbers of electrons from cloud 1 diffuse upstream and, secondly, 
that the electrons of cloud 2 ahead of cloud 1 have a uniform density and 
convect with the ions to preserve the quasi-neutrality of the upstream plasma. 
Figure~\ref{Fig8} demonstrates that no ions of cloud 1 leak upstream of its 
front and that no ions of cloud 2 are reflected at $x=300$ to form a beam 
upstream of cloud 1. Movies 1 and 2 together with Fig. \ref{Fig2} evidence, 
however, the growth of the EES and the onset of the filamentation upstream 
of cloud 1 prior to the time $T_1$. These two processes thus grow in an
interval, in which we only find the plasma of cloud 2 and no further particle
beams. We may thus exclude for our case study and until the time $T_{1}$ that 
particle beams (which can be formed by leaking downstream plasma 
or reflected fast ion beams in a quasi-parallel magnetic field geometry or in 
unmagnetized plasmas \citep{Malkov,Martins}), result in the growth of the EES 
and in the observed upstream instability.

Figure~\ref{Fig11} displays the fields at the time $T_{sim}$. The nonplanar 
$B_x$ reaches amplitudes comparable to $B_{x,0}/2$. Both components of 
$\vec{B}_\perp$ evidence planar oscillations with an amplitude exceeding 
$B_{x,0}$ and with a wavelength $\approx 120$. The wavelength is unchanged 
and the EES is still circularly polarized. The amplitudes of $B_y$ and $B_z$ 
peak in the interval $600<x<1000$ close to the front of cloud 1. Evanescent 
waves reach far upstream up to $x \approx 1700$. The $\vec{E}_\perp$ is more 
diffuse and $E_y$ shows structuring along $y$. Quasi-planar structures close 
to and ahead of the front of cloud 1 are revealed by $E_x$ for $700<x<1000$. 
The wave vectors of these oscillations point predominantly along $x$ and their 
characteristic wavelength is short compared to that of the EES. They resemble 
the structures of cloud 2 upstream of cloud 1 at late times in the movies 1 
and 2.  

The evolution of the fields is presented in the movie 3 for the three 
components of $\vec{B}$ and in the movie 4 for the three components of 
$\vec{E}$. Until the time $T_1$, when the overlap layer covers $|x|<300$, 
the fields can be subdivided into those in this layer and those of the EES. 

We discuss first the electromagnetic fields in the cloud overlap layer. 
The filamentation is observed best for $B_x$ and $E_y$. The $B_z$ is the
dominant component of the filamentation instability in our 2D geometry
and for a flow velocity vector along $x$ \citep{Bret,Sch1}. This component
indeed shows a filamentation, but the adaptation of the colour scale to 
the EES reduces the contrast and it is not clearly visible. The $B_x,E_y$
and $B_z$ show filamentary structures with a thickness below or equal to 
the ion skin depth. The filamentation of these field components shows 
the same flow patterns as the ion distribution in movie 1. The ion beam
filamentation is thus the primary driver of the fields. Oscillations of 
$E_x$ with a short wavelength along $x$ are observed, which we have related 
previously to a Buneman-type instability. 

The EES starts to grow immediately at the cloud collision boundary and 
it is thus a consequence of our initial conditions. The initial jump of 
the convection electric field $E_y$ across $x=0$, its constant amplitude 
along $y$ and the vanishing derivative along $z$ imply that ${\rm d} E_y 
/ dx = -\partial_t B_z$. A seed field for magnetowaves develops at the 
initial contact boundary. We may interpret its consequence as follows. 
The magnetic field is approximately stationary in the dense plasma. This
can be seen from the $E_y$ in the cloud overlap layer in Movie 4, which
is practically identical to that in cloud 1 outside the layer. The 
deflection of the electrons of cloud 2 by this $B_z$, which is localized 
in $x$, will be stronger than the deflection of the ions of cloud 2 and 
$J_y \neq 0$. The EES is unstable, which implies that the current amplifies 
the magnetic perturbation. The EES reaches a $B_\perp \approx B_{x,0}$ 
already when the cloud overlap layer has expanded to cover $|x|\la 100$. 
The EES is thus strong enough to cause the deflection of the ions at the 
tip of cloud 1 at the time $T_1$ in Fig.~\ref{Fig8}. The amplitudes of 
$B_y$ and 
$B_z$ grow and the EES expands steadily until secondary instabilities 
occur. 

When the cloud overlap layer covers $|x| \la 500$, the structures in $B_x$ 
upstream of cloud 1 in movie 3 start to accelerate along $y$. This 
acceleration is most easily visible close to the position, where the 
amplitude of the EES peaks. The 
time, when the structures of $B_x$ start to accelerate along $y$, coincides 
with a similar flow of the structures of $E_y$ and $E_z$ in movie 4. Prior to 
their acceleration, the structures in $B_x$ resemble the oblique structures 
in the plasma density of cloud 2 upstream of cloud 1 found at the time $T_1$ 
in the Figs.~\ref{Fig4} and~\ref{Fig5}. The density structures and those of 
$B_x$ are modulated along $x$ on scales comparable to the wavelength of the EES.

We may interpret the formation and the acceleration of the field structures 
in terms of an acceleration of the incoming particles of cloud 2 in the 
wave field of the EES. If its wave field is not perfectly uniform along $y$,
then the particle deflection is not equally efficient for all $y$, which
will cause density modulations. The particles of cloud 2 with their spatially 
nonuniform density are then accelerated in the $y,z$ plane by the EES, 
resulting in nonuniform currents. A spatially nonuniform current could drive 
the nonplanar field structures, which we observe in movies 3 and 4. A 
nonuniform $J_z$ would, for example, yield structures in $B_x,B_y$ and $E_z$ 
by $\partial_x B_y - \partial_y B_x = \partial_t E_z + J_z$. The motion of 
$B_x$ and also of $E_z$ and $B_y$ would then be induced by that of the 
density structures. The growth and the accelerating motion along $y$ of the 
structures are presumably tied to the increasing amplitude of the EES.

Movie 4 shows that the $E_x$ oscillates outside the cloud overlap layer. 
The amplitude of $E_x$ is constant within each cloud on scales much larger 
that $\lambda_s$ and the oscillations start instantly. Thus, they cannot 
be connected to the processes at $x \approx 0$, because of the finite light 
speed. These oscillations are periodic in time and their frequency is the 
electron plasma frequency of the respective cloud. These oscillations 
presumably originate from a slight initial mismatch of the currents of 
electrons and ions along $x$ in each cloud, which results through 
$\partial_t E_x = -J_x$ in electrostatic oscillations at the local plasma 
frequency. 

Structures in the plasma of cloud 2 with a wave vector that is aligned with 
$x$ were observed in the Figs.~\ref{Fig9} and~\ref{Fig10} for $750 < x < 
1200$. Movies 3 and 4 provide additional information. These short-scale
structures do not show up in $\vec{B}_\perp$ and hence they do not have a 
strong electromagnetic component. The modulations are most clearly visible 
from $E_x$, for example in Fig.~\ref{Fig11}(d). A polarization along 
$\vec{v}_b$ indicates that these waves are electrostatic. The density 
structures in cloud 2 should thus be charge density waves, similar to the 
Buneman waves we have observed in Fig. \ref{Fig2}(d). This charge density
wave is not the only one upstream of cloud 1. Movie 3 shows at late times 
and for large $x>0$, that magnetowaves with a longer wavelength than that 
of the EES are generated that move faster than the EES. 

\begin{figure}
\centering
\includegraphics[width=\columnwidth]{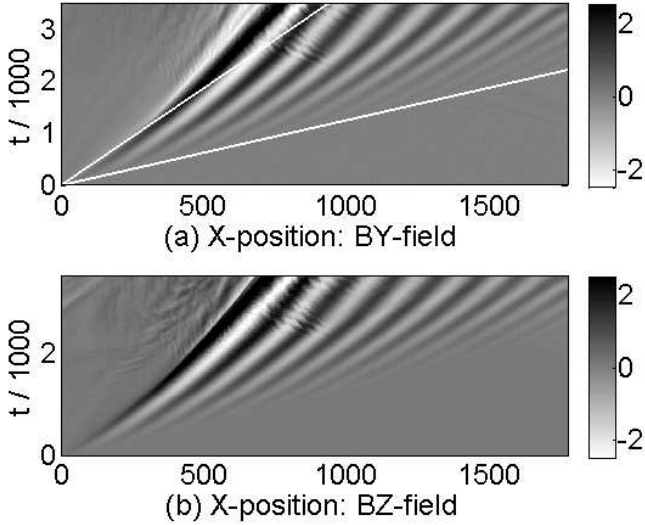}
\caption{The $\vec{B}_\perp (x,y=0,t)$ in the reference frame of the
simulation box. The $B_y$ is shown in (a). The two white lines correspond 
to $x = v_b t$ and $x=0.8 ct$. The $B_z$ is shown in (b).}\label{Fig12}
\end{figure}

A wave conversion in form of a four-wave coupling of a circularly polarized, 
large amplitude wave has been proposed by \citet{Instability}. The four 
interacting waves all have parallel wave vectors and they will thus all 
result in plasma and field modulations along $x$ in our geometry. The 
four-wave interaction results in a forward-propagating magnetowave with
the same polarization as the driving wave, which might be the long-wavelength 
magnetowave observed in movie 3. It is faster than the super-Alfv\'enic EES,
implying that it should propagate in the high-frequency branch of the 
left-hand polarized mode \citep{Treumann}. This parametric wave interaction 
also pumps a charge density wave, which could result in the density modulation 
observed in the movies 1 and 2 ahead of cloud 1. Finally, the interaction 
should produce another magnetowave that moves in the opposite direction than 
the EES.

We could not identify this fourth wave in $\vec{E}(x,y,t_0)$ and 
$\vec{B}(x,y,t_0)$ at any given time $t_0$. This either means it is not
there or that its amplitude is close to noise levels. Its wave vector is 
aligned with that of the EES and it is a magnetowave \citep{Instability}. 
It should thus be detectable in slices of $\vec{B}_\perp (x)$ for $y=0$, 
which we can plot as a function of time. The space-time correlation allows 
us to identify structures, which are not easily seen from individual time 
slices. Figure~\ref{Fig12} shows them as a function of time. 
Both components of $\vec{B}_\perp$ show at $x\approx 800$ and $t\approx 2700$ 
waves propagating antiparallel to the EES. These fast waves originate from 
processes taking place at the rear end of cloud 2, which are fuelled by the 
different expansion speeds of electrons and ions into a vacuum. These
waves can be seen already before they start to interact with the EES.
The $B_y$ amplitude is modulated at $x\approx 1700$ and $t\approx 1100$. 
These waves are amplified as they pass through the EES. We have to emphasize 
though, that the backward-propagating magnetowave is only barely visible in 
this plot and that we have not shown that its amplification by the EES 
results from a coherent 4-wave interaction. The latter would require more 
detailed correlation studies of the wave's phases. Our current simulation
data is too noisy and we leave this to dedicated 1D PIC studies that can
employ a higher number of particles per cell to reduce noise levels.

Two lines are fitted in Fig.~\ref{Fig12}(a). The $x=v_b t$ is approximately 
twice the Alfv\'en speed $V_A$ of the cloud 2. Initially, the rear end of the 
EES moves with $v_b$ and it is thus connected to the front of cloud 1. The 
speed of the EES decreases steadily and its rear end at the time $T_{sim}$ 
coincides with the front of cloud 1 at $x\approx 750$ rather than $v_b T_{sim} 
\approx 950$. This deceleration makes it difficult to determine the frequency 
of the EES, which should be measured in its rest frame. The first maximum of 
$B_y$ moves with the front of cloud 1 and shows no change in time. We may 
conclude that its oscillation frequency is low or zero in this decelerating 
reference frame. The $B_z$ shows a change at the front of cloud 1, which may 
evidence a frequency $\omega_{EES} \neq 0$ of the EES. It may, however also 
be related to the formation of the shock. We can thus only conclude here that 
$\omega_{EES} \la 2\pi / T_{sim} \approx (m_e/m_i)\omega_{ce}$. The second 
fitted line in Fig.~\ref{Fig12}(a) corresponds to $x = 0.8ct$ and it is an 
approximate boundary for the front of the EES. This expansion speed is 
$0.87c$ in the reference of cloud 2.

\begin{figure}
\centering
\includegraphics[width=\columnwidth]{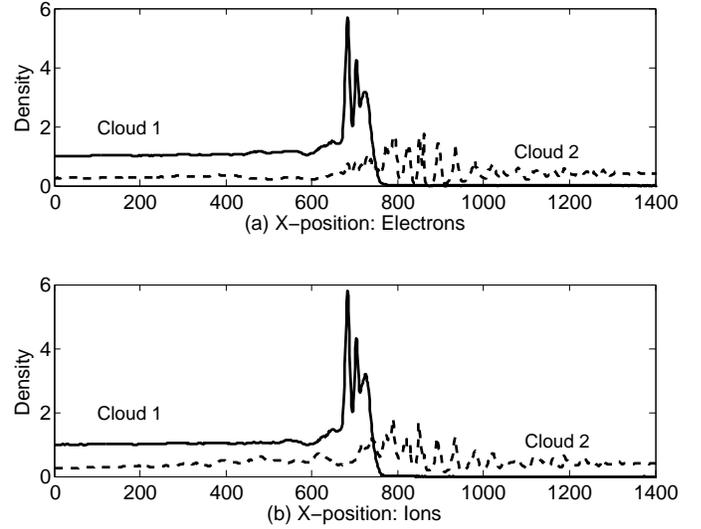}
\caption{Densities integrated over $y$ at the time $T_{sim}$. (a) shows the 
electron densities and (b) the ion densities. The solid curve corresponds 
to cloud 1 and the dashed curve to cloud 2. The latter is multiplied by the 
factor 4.}\label{Fig13}
\end{figure}

Figure~\ref{Fig13} shows the densities of both clouds, averaged over $y$, 
at the time $T_{sim}$. We consider only $x>0$. The densities of all four 
species are practically constant for $x<600$ and the density of each species 
is comparable to its initial value. The processes in the cloud overlap layer 
have reached a state that does not involve a density modulation. No shock 
is present in this interval, because that would result in a plasma compression.
The plasma density of cloud 1 rises in the interval $600 < x < 750$ and 
decreases to zero for $x>800$. The front of cloud 1 is thus still well-defined. 
The density of cloud 1 increases to about 3-4 times its initial value in the 
interval $650<x<750$, reaching a peak density of 6. The plasma density of 
cloud 2 reveals substantial oscillations for $x>800$ upstream of cloud 1. 
The electron and ion density oscillations, which could be observed in the 
movies 1 and 2, and the $y$-averaged densities do not match perfectly. The 
resulting net charge modulation results in charge density waves and the 
$E_x \neq 0$ in Fig.~\ref{Fig11}(d).

\begin{figure}
\centering
\includegraphics[width=\columnwidth]{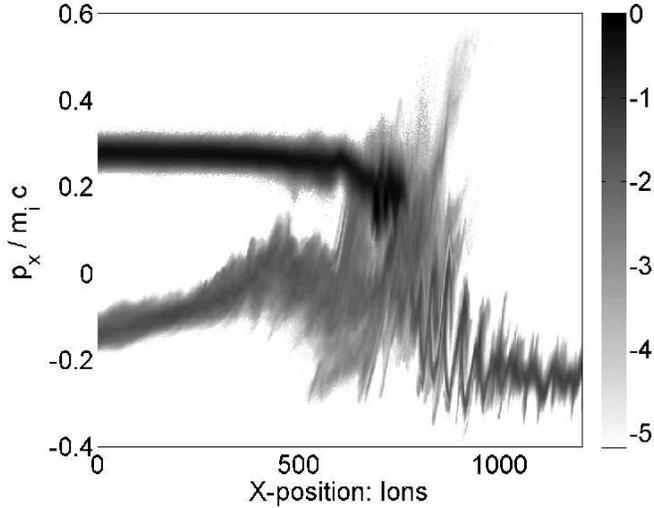}
\caption{Ion phase space density at the time $T_{sim}$. The grayscale shows the 
10-logarithmic density normalized to the density peak.}
\label{Fig14}
\end{figure}

The ion phase space density distribution in the $x,p_x$ plane and at the
time $T_{sim}$, which has been integrated over all other position and momentum 
components, is shown in Fig.~\ref{Fig14}. 
The ions of both clouds have started to mix in the $x,p_x$-space, which is 
the condition for a shock formation. The ion clouds merge in the interval 
$600 < x < 800$, which coincides with the interval that has an increased 
plasma density in the Fig.~\ref{Fig13}. The density increase is thus due 
to the shock compression and it will result in a downstream region that 
expands along $x$. This future downstream plasma has not yet thermalized, 
explaining why the density in Fig.~\ref{Fig13} is not constant within 
$600 < x < 800$. 

The forward shock, which will be moving to higher $x$, will be somewhat 
slower than the collision speed of $c/2$. It will move faster than $v_b$ 
in the reference frame of cloud 2. This is because the plasma has a net 
momentum in the simulation frame due to the higher density of cloud 1. 
The shock moving in the direction of decreasing $x$ will move slower than 
$v_b$ in the reference frame of cloud 1. We can already see the slowdown 
of the front of cloud 1, which coincides with the amplitude maximum of
the EES in Fig.~\ref{Fig12}, which will form the forward moving shock. 

Figure~\ref{Fig14} also reveals oscillations of the mean momentum $\langle 
p_x \rangle (x)$ of cloud 2 in the interval $800 < x < 1000$. The waves
that can cause such strong oscillations must yield clearly detectable field 
oscillations in the same interval. They must be caused either by the charge 
density waves and the associated $E_x$ or by the EES ahead of cloud 1, or 
by the combination of both. The wavelength of the oscillations is well below 
the wavelength $\approx 120$ of the EES and comparable to that of the 
oscillations of $E_x$ in Fig.~\ref{Fig11}(d) with a wavelength $\lambda_W/3$. 
The electrostatic potential of the charge density waves will indeed cause 
particle oscillations in the $x,p_x$ plane. The main effect of the EES, 
namely the gyro-bunching, is not necessarily visible from the phase space 
density distribution in the $x,p_x$ plane. The $B_\perp$ of the EES forces 
the ions on a gyro-orbit orthogonal to $x$. The electrostatic component of
the EES provided by $E_x$ will, however, modulate the ion distribution in 
the $x,p_x$-plane. No clear modulation of the ion paths on scales $\sim
\lambda_W$ is visible in Fig.~\ref{Fig14} though. 

\begin{figure}
\centering
\includegraphics[width=\columnwidth]{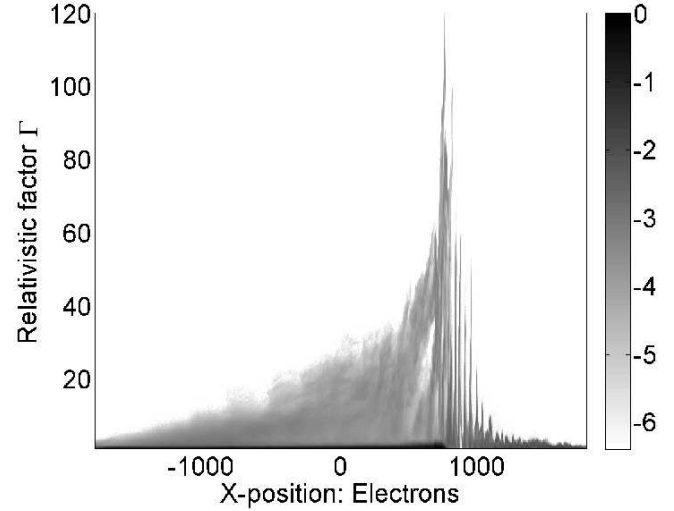}
\caption{Electron phase space density at the time $T_{sim}$. The grayscale 
is the 10-logarithmic density normalized to its peak.}
\label{Fig15}
\end{figure}

The electrons are accelerated to highly relativistic speeds in the interval
in which we find the charge density waves and the forming shock. The peak 
Lorentz factor $\gamma$ is 120, according to the Fig.~\ref{Fig15}.
This peak Lorentz factor is remarkable in that it exceeds even the $(m_i/m_e) 
(\gamma [v_c] -1) \approx 60$, which we would obtain if the energy of an 
electron would equal that of an ion moving with the cloud collision speed 
$v_c = c/2$. This energy is, however, reached only by a tenuous subpopulation 
of the electrons in a limited spatial interval. This strong electron 
acceleration is presumably achieved by those fields, that give rise to 
what will become the shock-reflected ion beam. The ions of cloud 2 at 
$x\approx 750$ in Fig.~\ref{Fig14} are accelerated from the speed $-v_b$
to the speed $0.5c>v_b$ at this position and their velocity thus changes by
more than $v_c$. We also observe that the electrons are not only accelerated 
in the forming downstream region $600 < x < 800$, but they appear to be 
pre-accelerated in the interval $800<x<1100$. The electron distribution 
shows narrow spikes in this region. The electron phase space density is 
smooth in the interval $500 < x < 750$ and the energies are here comparable 
to the energy of an ion moving with $v_c$. The peak energy of the electrons 
decreases steadily, as we go from $x\approx 500$ to $x\approx -1500$. The 
characteristic distances crossed by electrons with $\gamma > 5$ will 
not differ much during the short time $T_{sim}$. This distance is thus 
proportional to the time, which has passed since they were accelerated. The 
energy to which electrons have been accelerated must therefore have increased with time.

\begin{figure}
\centering
\includegraphics[width=\columnwidth]{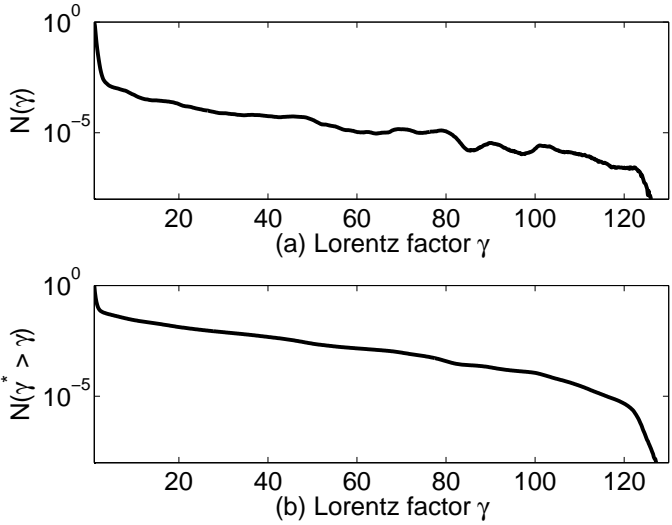}
\caption{Electron energy distributions, sampled at the time $T_{sim}$ and
normalized to their respective peak value, as a function of $\gamma$. The 
number density $N(\gamma)$ is displayed in (a), while (b) shows 
$\tilde{N}(\gamma) \equiv N(\gamma^* > \gamma) = \int_\gamma^\infty N(\gamma^*) 
d\gamma^*$.}\label{Fig16}
\end{figure}

Figure~\ref{Fig16} displays the electron phase space density from 
Fig.~\ref{Fig15}, which has been integrated over all $x$ to give 
$N(\gamma)$. It also shows the probability $\tilde{N}(\gamma) = 
\int_\gamma^\infty N(\gamma^*) d\gamma^*$ of finding electrons with a Lorentz 
factor $\gamma^* > \gamma$.
Most electrons have a $\gamma < 5$. The density of this cool population is 
exaggerated, because we integrate over the full simulation box and, thus, 
over regions which have not yet been affected by the cloud collision. The 
high energy tail in the interval $5 < \gamma < 100$ in both distributions 
can be approximated by an exponential and the densities decrease rapidly to 
zero for $\gamma > 120$. 

Movie 5 provides further insight into the electron acceleration mechanisms. 
It is a time-animation of the phase space density distributions $f_i (x,p_x)$ 
of the ions and $f_e (x,\gamma)$ of the electrons until the time $T_{sim}$. 
The colour scale denotes the 10-logarithmic number of computational particles 
in units of those of cloud 2. One particle of cloud 1 corresponds to 10 of
cloud 2.

During the initial phase of the cloud collision, up to the time when the 
ions of both clouds overlap in the interval $|x| \la 50$ in movie 5, the 
ions propagate practically freely. The ion phase space distribution at 
the front of each cloud is increasingly tilted, because ions with a higher 
speed propagate farther in a given time interval. The electrons are moderately
heated to $\gamma \approx 2.5$ at the front of cloud 2. The heating mechanisms 
are the filamentation instability and the Buneman-type instability, which also 
increase the thermal spread of the ions of cloud 2 at its tip in 
Fig.~\ref{Fig8}(b). 

When the cloud overlap layer covers $|x| \la 300$, the electron heating 
at the front of cloud 2 has ceased and the electron acceleration sets in
at the front of cloud 1. The slowdown of the ions of cloud 1 is visible
and it increases steadily, which is consistent with a magnetic deflection
by the growing EES. When the ions of cloud 1 have reached $x\approx 400$, 
the electrons are accelerated to $\gamma \approx 6$. The movie 5 shows that 
this electron acceleration is limited to the interval covered by the EES. 
The electrons of cloud 2 are accelerated as they approach the front of
cloud 1. This acceleration results in a beam in $f_e (x,\gamma)$ prior to 
their encounter with the front of cloud 1 and a depletion of nonrelativistic 
electrons. The movie 5 shows this as a detaching of the electron distribution 
from the coordinate axis $\gamma = 1$. 

We may interpret this as follows. The electrons of cloud 2 are dragged with 
the ions across the $\vec{B}_\perp$ of the EES and all electrons experience 
the same accelerating force during the same time interval, resulting in a 
similar velocity increment for all electrons. The initial electron beam 
stays compact and the electrons move on a corkscrew orbit in the circularly 
polarized EES \citep{Shock1}. After this acceleration, the electrons of 
cloud 2 cross the front of cloud 1 and they are mixed in phase space to 
form a smooth density distribution. Movie 5 demonstrates that the electron 
acceleration by the EES increases steadily in time, presumably because their 
cross-field transport across the EES with its increasing amplitude and extent 
along $x$ provides a stronger acceleration for a longer time.

\begin{figure}
\centering
\includegraphics[width=\columnwidth]{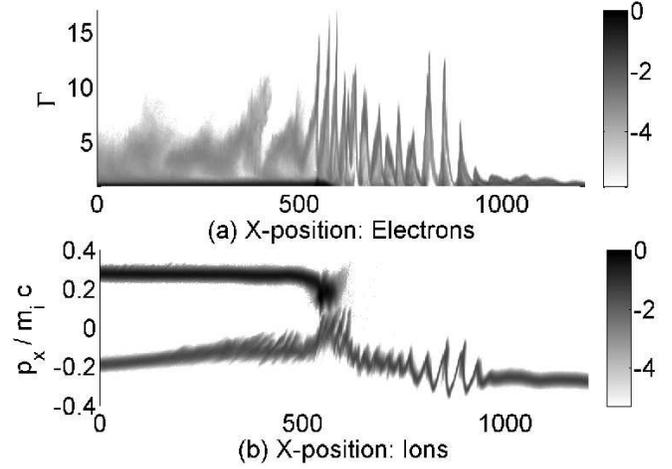}
\caption{Phase space distributions at the time $0.8 T_{sim}$. (a) 
displays the electrons and (b) the ions.}\label{Fig17}
\end{figure}

When the ions of cloud 1 have propagated to $x\approx 500$ in movie 5, the 
ions of both clouds start to merge. We show the phase space densities 
$f_i (x,p_x)$ and $f_e (x,\gamma)$ of both species directly after this 
merger at the time $0.8T_{sim} \approx 2800$ in Fig.~\ref{Fig17}. 
We notice at this time localized and strong oscillations of $\langle p_x 
\rangle (x)$ at $800 < x < 1000$ and weaker ones with a shorter wavelength
within the interval $400 < x < 800$. The electrons are accelerated to highly 
relativistic energies by these structures, as expected. Consider a 
force that causes velocity oscillations of the ions of $\Delta V \approx 
0.1c$ in Fig.~\ref{Fig17}(b). The same force will clearly result in 
relativistic electron velocity oscillations. The oscillations of the ion 
mean momentum move in the negative $x$-direction in movie 5 and they are 
amplified as they pass through the EES. 

We may associate these wave structures with the backward propagating charge 
density waves and magnetowaves expected from the four-wave interaction, 
although we have to point out that the strong plasma modulations imply a 
nonlinear stage of the plasma, which is not incorporated into the equations 
describing the four-wave interaction. A consequence can be that e.g. the 
strong and compact magnetowave modulates the plasma density by the 
ponderomotive force of its fields. The fields resulting in the large 
modulation must belong to a well-defined wave packet. Figure~\ref{Fig12} 
evidences at the time $0.8T_{sim}$ the presence of a magnetowave in the 
interval $800 < x < 1000$, in which we find the strong momentum oscillation 
of cloud 2 in Fig.~\ref{Fig17}(b). The oscillations of $\langle p_x \rangle 
(x)$ in the interval $600 < x < 700$ may correspond to the charge density 
wave, which we observed at late times in the movies 1 and 2. Both, the charge 
density waves and the magnetowave accelerate electrons to highly relativistic 
speeds.

\begin{figure}
\centering
\includegraphics[width=\columnwidth]{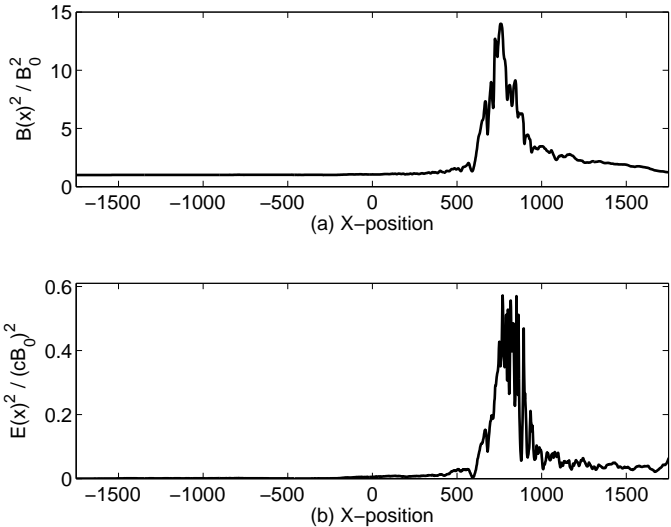}
\caption{Field energy densities at the time $T_{sim}$. (a) considers the 
magnetic energy density and (b) that of the electric field.}\label{Fig18}
\end{figure}

We finally assess the energy contained by the fields at the time $T_{sim}$ 
and the value of plasma collisions of the type we have considered here for 
the magnetic field amplification upstream of the main blast shell of a SNR. 
The magnetic energy density in physical units $\vec{B}^2 / 2\mu_0$ is 
normalized to $\vec{B}_0^2 / 2\mu_0$ and the electric energy density in 
physical units $\epsilon_0 \vec{E}^2/2$ is normalized to $c^2\vec{B}_0^2 / 
2\mu_0$. Both are plotted in Fig.~\ref{Fig18}. Significant field energy 
densities are found only close to the front of cloud 1 at $x\approx 750$. 
The magnetic field dominates and its peak energy density exceeds that of 
the electric field by the factor $\approx 25$. The slowdown of the front 
of cloud 1 implies that the $\vec{E}_\perp = -\vec{v}_f \times \vec{B}_\perp$ 
decreases, where $\vec{v}_f$ is the speed of the front of cloud 1 that was 
initially $\vec{v}_b$. The energy density of the magnetic field is amplified 
by over an order of magnitude by the plasma collision and by the shock 
formation. The strongest magnetic fields corresponding to the circularly 
polarized EES coincide with the fastest electrons in the simulation. This 
interval should emit significant electromagnetic radiation.

\section{Discussion}

We have described in this paper the collision of two plasma clouds at the
speed c/2. The ion to electron mass ratio of 400 has allowed us to model 
the collision in two spatial dimensions until the shock forms. Then we had 
to stop the simulation. The acceleration of electrons to speeds $\sim c$ 
and the rapid expansion speed of the energetic electromagnetic structure 
(EES) imply, that both will quickly reach the boundaries; hence our periodic 
boundary conditions become invalid. Open boundaries would allow the electrons 
and the wave energy to flow out of the system. However, the instabilities 
driven by these beams \citep{Martins} and by the EES are important for the 
upstream dynamics and the latter will be adversely affected by open boundaries. 

The filamentary structures can form and merge in the 2D geometry we consider
here up to the instant when the magnetic repulsion of two filaments with 
oppositely directed current enforces their spatial separation 
\citep{Davidson,Lee}, at least in the absence of an oblique magnetic field. 
Then no further mergers occur since only one dimension is available 
orthogonal to the flow velocity vector. A realistic 3D PIC simulation would 
allow the filaments to move around each other and merge with other filaments 
of equal polarity \citep{Lee}. A 3D PIC simulation is, however, currently 
impossible for our case study involving ions, because of the computational 
cost involved in resolving ion and electron scales. Simulations in three 
spatial dimensions are now feasible for the case of leptonic shocks 
\citep{Nishi2}.

The densities of the clouds differ by a factor of 10 and the collision 
is asymmetric, as in the simulation by \citet{Oblique1,Oblique2,Sorasio}. 
Initially the magnetic field is uniform and quasi-parallel to the flow 
velocity vector. Its significant strength together with the high plasma 
temperature of 25 keV and the unequal cloud densities reduce the growth 
rate of the filamentary instabilities \citep{Mag,Bellido}. 

\citet{Shock1} have previously probed the higher $\gamma$ regime appropriate 
for the internal shocks of gamma-ray bursts. Here we consider the mildly 
relativistic regime, with a collision speed $0.5c$ between both clouds. 
Such a speed might be realistic for a plasma subshell outrunning the main SNR 
shock. Such subshells can reach relativistic flow speeds for particularly 
violent SNR explosions \citep{Shockspeed1}. In our initial conditions, the 
magnetic energy surpasses  the thermal energy of the dense plasma slab by 
a factor of 5. The plasma flow implies, however, that the box-averaged 
plasma kinetic energy density exceeds the magnetic energy density by an order 
of magnitude. We summarize several aspects of our results.

\subsection{Effects due to initial conditions}

Our initial conditions have resulted in the formation of planar wave and
plasma structures, the most important one being the EES. We think that the 
EES grew out of a localized seed magnetic field pulse driven by the spatial 
gradient of the convection electric field. The plasma upstream of the dense 
cloud is destabilized by this electromagnetic structure and the EES expands 
at the speed $0.87c$ in the reference frame of the tenuous cloud. The energy 
for its growth and expansion is provided by the kinetic energy of the 
upstream medium, which moves with respect to the EES. The shock speed 
$\la v_c$ then implies that we have a coherent magnetic layer that expands 
its width at a speed of at least $(0.87c-v_c)/(1-0.87 v_c / c) \approx 0.65c$, 
measured in the reference frame of the tenuous cloud. It covered about 80 ion 
skin depths at the end of the simulation, showing no signs of a slowdown. 

The EES is a consequence of our initial conditions and the growth of the 
seed magnetic field amplitude could probably be reduced but not suppressed 
by a smoother change of the convection electric field, which can be achieved 
by a gradual change of the plasma convection speed \citep{Oblique1,Oblique2}. 
However, the seed magnetic field could be provided also by waves with a short 
wavelength, e.g. whistlers, and it is thus not unphysical. 

Structures with strong magnetic fields, similar to the EES and known as SLAMS, 
are frequently observed close to quasi-parallel shocks in the solar system 
plasma and they can accelerate electrons to high energies. They are thus 
potentially important also for SNR shock physics. 
Our initial conditions provide a possibility to let nonlinear MHD waves grow 
out of a simple simulation setup for a further study. The EES is moving 
with the ions of the dense cloud and it modulates the ions and electrons 
of the tenuous cloud, thereby gaining energy. Its growth probably requires 
an asymmetric plasma collision. 

\subsection{Shock formation}

In this paper we modelled the formation of the shock from the initial 
collision of two plasma clouds. The signatures of the shock are evident, 
including visible thermal broadening behind the shock and a dense shock ramp. 
While filamentation structures form ahead of and behind the shock, we note 
that the structure is basically planar in the critical foreshock area, where 
electron acceleration is expected to occur. This means that one-dimensional 
simulations will be relevant in this region, allowing much higher resolution, 
increased particle number (resulting in lower particle noise and a better
phase space resolution) and a higher ion-electron mass ratio, than is 
currently found in two and three-dimensional simulations. 

It is evident from the simulation that the filamentation is not fully 
suppressed by the guiding magnetic field and by the high plasma temperature. 
Its amplitude has been set such that it should suppress the electron 
filamentation if the plasma would be spatially uniform \citep{Mag}. This 
amplitude is apparently insufficient to suppress the slower filamentation 
of the ion beams and we could even see evidence for an electron beam 
filamentation just behind the front. The front of the dense plasma cloud 
maintains its planarity throughout the simulation, but even here the density 
was non-uniform along the boundary. The onset of the filamentation was, 
however, delayed. The likely cause is the high density gradient across the 
front, which alters the electron and ion skin depths and thus the 
characteristic scale of the filaments. The gradient is caused by the slowdown 
of the ions by the magnetic field of the EES and by the electron acceleration. 
The electrons are confined at the front in the direction of the shock normal
so that they preserve the quasi-neutrality of the plasma, but they can move
orthogonally to it. The latter results in a drift current. 

\subsection{Field amplification}

\citet{MagAmp2} and \citet{MagAmp1} have described observations of magnetic 
field amplification above the value expected from shock compression in 
SNRs. At the final simulation time the magnetic field energy 
density is increased in strength by over one order of magnitude, exceeding
by far that expected from the magnetic field compression by the shock. A 
shock compresses only the magnetic field component perpendicular to the 
shock normal, which is weak in our case, and the amplification of its energy
density can reach a factor of 4-7. The magnetic energy density at the 
simulation's end has been comparable to the box-averaged kinetic energy 
density in an interval spanning about 10-20 ion skin depths. The magnetic 
energy density due to the EES was at least twice as high as that of the 
background field in an interval covering 50 ion skin depths. Even if we take 
into account that the kinetic energy density close to the shock is increased 
by the accumulation of plasma, the magnetic energy density still constitutes 
a sizeable fraction of the local total plasma kinetic energy density. The EES 
has provided the main contribution to the magnetic energy density and exceeded 
that due to the filaments downstream by two orders of magnitude. 

Throughout this paper, we used normalized quantities in our simulation and 
we can scale the magnetic field amplitude to the relevant plasma conditions. 
If we set the electron density of the dense cloud to 1 cm$^{-3}$, we would 
obtain a peak magnetic field with a strength of 10 mG. However, we have
to point out that our initial magnetic field amplitude has been higher than 
that expected for the ambient plasma, even if we take into account its
amplification by cosmic ray-driven instabilities, and our simulation results
may not be directly applicable.

Where does the extra field come from? Amplification of the magnetic field 
can occur from the electron drift current arising from the $\vec{E} \times 
\vec{B}$ drift motion in a layer close to the shock that is narrower than 
the ion gyroradius but wider than the electron gyroradius, see, e.g. 
\citet{BaumjohannTreumann}. The current adds to the shock current and 
increases the jump in the perpendicular magnetic field. This can only 
occur when the ion and electron gyroradii differ, i.e. not in a pair plasma. 
The EES has a significant $E_x$-component and $|\vec{B}_\perp| \approx 
|\vec{B}_0|$. We thus obtain a $\vec{E}\times \vec{B}$ drift orthogonal to 
the flow velocity vector.

We have also found that the requirement to maintain quasi-neutrality of the
plasma implies that the upstream electrons are dragged with the upstream
ions across the EES, which moves with the shock. The resulting 
$\vec{v}\times\vec{B}$ drift accelerates the electrons orthogonally to
the shock propagation direction, further enhancing the net current and
the magnetic field. Finally, magnetic fields of SLAMS are provided by the 
current due to the gyro-bunched ions, which rotate in the plane orthogonal 
to the wavevector.

These mechanisms increase the mean magnetic field, and are different from 
the instability described by \citet{MagAmp3} who has described a cosmic ray 
streaming instability which can amplify turbulent magnetic fields ahead of 
the shock. We can exclude Bell's instability here since we have not found 
energetic particles with a significant density moving upstream, which would 
provide the net current that drives this instability.

\subsection{Electron acceleration and upstream wave spectrum}

The shock retains its planar structure after it forms. A circularly polarised 
large-scale precursor wave, the EES, expands into the foreshock. Its 
wavelength is several times the ion skin depth. It gradually rotates the 
quasi-parallel magnetic field into a quasi-perpendicular one at the current 
layer and it forces the incoming ions and electrons to interact with it
nonlinearly. The ions are gyro-bunched and some of the incoming ions of
the tenuous cloud are reflected by the forming shock. We have found evidence 
of a parametric instability \citep{Instability} of the EES ahead of the 
foreshock and we could find at least two waves that may be the result of this 
parametric decay. These waves appear at late times, when the EES has expanded 
in space and is thus sufficiently monochromatic. They grow to an amplitude 
that introduces oscillations of the mean velocity of the ions of up to $c/5$. 

The interplay of the short-scale charge density waves and magnetowaves causes 
the electrons to be accelerated to highly relativistic speeds upstream of the 
forming shock. Similar electron acceleration (injection) mechanisms upstream 
of shocks involving whistler waves have been proposed by 
\citet{Oblique3,Injection3,Gyrosurf}. The strongest electron acceleration is, 
however, observed at the location where the shock-reflected ion beam is 
developing. The electrons are accelerated to a peak Lorentz factor of 120 and 
their energy gain is thus comparable to the energy associated with the 
velocity change of the shock-reflected ions. If the electron acceleration is 
accomplished by the electromagnetic fields that reflect the incoming upstream 
ions, then the energy gain of the electrons may scale with the ion mass. We 
may expect in this case that the electrons are accelerated to a Lorentz factor 
$\gamma_M \approx 120 m_p / m_i$ that is $\gamma_M \approx 550$ if we would
use the correct proton to electron mass ratio.

It is interesting to see if this type of electron acceleration can also occur 
close to Solar system shocks. Let us consider the Earth bow shock as one of 
the best known collisionless shocks and let us assume a Solar wind speed of 
$4 \times 10^5$ m/s to $7.5 \times 10^5$ m/s. A specular reflection of the 
incoming Solar wind protons by the bow shock would change their energy by 
about 0.8 keV to 3 keV. Electrons with such energies are observed in a thin 
sheet close to the shock surface of perpendicular shocks \citep{BowShock}. 
Perpendicular shocks are capable to produce shock-reflected ion beams
\citep{Shock6} and the electron acceleration mechanism we observe here
may work also at the Earth bow shock. 
   
\subsection{Future work}

This simulation study was concerned with the collision of two plasma clouds 
at a mildly relativistic speed. Its purpose has been to better understand
the conditions and the mechanisms involved in the formation of a shock.
This shock will move at an essentially nonrelativistic speed below the 
initial collision speed and, thus, be relevant for fast SNR flows. A 
two-dimensional simulation geometry was necessary to assess the importance
of the multi-dimensional filamentation instability for the shock dynamics.
A quasi-parallel guiding magnetic field was used to slow down this
filamentation, resulting in a planar (one-dimensional) shock. 

The formation of the shock could be observed, but the simulation had to
be stopped at this time due to computational constraints. This
simulation has, however, revealed several aspects that should be examined
in more detail in more specialised simulation studies. 

The EES probably formed in response to our initial conditions. i.e. the
sharp jump in the convection electric field at the cloud collision boundary.
It has to be investigated if the EES also forms if this jump is decreased,
for example by a higher-order field interpolation scheme or by a gradual
decrease of the convection electric field by a smooth change in the plasma
convection speed. The simulations by \citet{Oblique1,Oblique2} suggest
that this will leave unchanged the magnetic field amplification and the
electron acceleration. However, the EES may not be so strong and coherent.

The magnetic field amplitude in the present study is higher than it is 
realistic for a SNR scenario. Future studies must address how the shock 
formation depends on lower amplitudes of the guiding magnetic field. 
Computationally inexpensive parametric simulation studies that resolve 
only one spatial dimension may provide insight. Initial studies not discussed 
here indicate that the shock formation is delayed by a decreasing magnetic 
field amplitude.

It is also necessary to follow the plasma collision for a longer time. An
important aspect is here how far the EES can expand upstream and how
strong the downstream magnetic field is. The planarity of the EES and of
the shock boundary may permit us to use one-dimensional simulations, by
which we can expand by at least an order of magnitude the box size along 
the collision direction. 

An one-dimensional simulation also allows us to examine with a larger number 
of particles per cell and, thus, lower noise levels the secondary instabilities
driven by the EES. We have found evidence for an instability of the EES to 
a four-wave interaction. Lower noise levels would allow us to compare the 
amplitudes and phases of the EES with those of the secondary waves, which
is necessary to demonstrate a coherent wave interaction. Extending the 
simulation time together with suitable initial conditions would, potentially, 
allow us to investigate what happens if the speed of the EES decreases below 
the local Alfv\'en speed. It is possible that the EES decouples from the shock 
and propagates independently in form of an Alfv\'en wave packet.

Finally, it would be interesting to reduce the collision speed to about
$c/10$, which is close to the expansion speed of the SNR shock, to see 
if and how many electrons are accelerated to relativistic speeds.
This will provide insight into the electron injection efficiency of oblique 
shocks and, thus, into the ability of SNR shocks to accelerate electrons to 
cosmic ray energies. 

\begin{acknowledgements}
GCM is funded by the Science Foundation Ireland grant 08/RFP/PHY1694. MED 
is funded by Vetenskapsr\aa det and by the Deutsche Forschungsgemeinschaft 
through the grant FOR1048. This research was supported in part by the 
National Science Foundation under Grant No. PHY05-51164. The authors thank 
Prof. M. Pohl and Dr. J. Niemiec for useful discussions related to the 
initial conditions. The Swedish HPC2N and the Irish ICHEC have provided the 
computer time and support. 
\end{acknowledgements}


\end{document}